# Molecular dynamics and charge transport in highly conductive Polymeric Ionic Liquids

Falk Frenzel,*,† Ryan Guterman,‡ A. Markus Anton,† Jiayin Yuan,‡ and Friedrich Kremer†

*Leipzig University, Institute of Experimental Physics I, Linnéstrasse 5, 04103 Leipzig, Germany, and Max Planck Institute of Colloids and Interfaces, Department of Colloid Chemistry, Am Mühlenberg 1 OT Golm, 14424 Potsdam, Germany*

E-mail: falk.frenzel@physik.uni-leipzig.de

**Abstract**

Glassy dynamics and charge transport are studied for the polymeric Ionic Liquid (PIL) poly[tris(2-(2-methoxyethoxy)ethyl)ammonium acryloxypropyl-sulfonate] (PAAPS) with varying molecular weight (9700, 44200, 51600 and 99500 g/mol) by Broadband Dielectric Spectroscopy (BDS) in a wide frequency ($10^{-2}$ - $10^{7}$ Hz) and temperature range (100 - 400 K) and by DSC- and AC-chip calorimetry. The dielectric spectra are characterized by a superposition of (i) relaxation processes, (ii) charge transport and (iii) electrode polarization. The relaxation processes (i) are assigned to the dynamic glass transition and a secondary relaxation. Charge transport (ii) can be described by the random free-energy barrier model as worked out by Dyre et al.; the Barton-Namikawa-Nakajima (BNN) relationship is well fulfilled over more than 8 decades. Electrode polarization (iii) follows the characteristics as analyzed by Serghei et al.;

*To whom correspondence should be addressed

†Leipzig University, Institute of Experimental Physics I, Linnéstrasse 5, 04103 Leipzig, Germany

‡Max Planck Institute of Colloids and Interfaces, Department of Colloid Chemistry, Am Mühlenberg 1 OT Golm, 14424 Potsdam, Germany

with deviations on the low frequency side. The proportionality between the relaxation rate of the dynamic glass transition and the charge carrier hopping rate reflects the nature of charge transport as glass transition assisted hopping. Hereby, the PIL under study exposes the highest dc-conductivity values observed for this class of materials below 100 °C, so far; and for the first time a conductivity increase by rising degree of polymerization. The comparison of the polymeric Ionic Liquids under study with others implies conclusions on the design of novel highly conductive PILs.

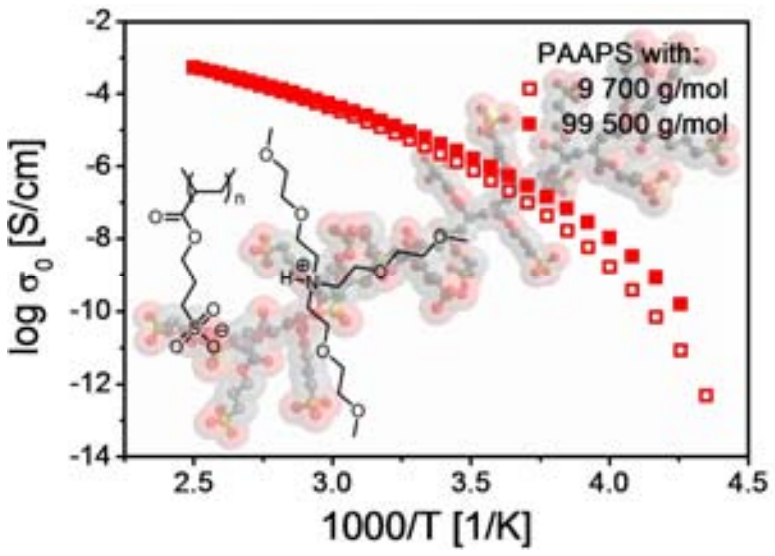

Figure 1: TOC-figure

# Introduction

Since the 1970s, when the first time a lithium battery has been reported their continuous evolution to consistently smaller, and even more fractal structures enhanced the storage capacity up to 300 mA h g$^{-1}$ [1] making them the technology of choice for most electrical storage applications covering the range from mobile device batteries up to offshore wind parks. It's surpassing power density is mainly a result of the small size of lithium ions and multiple combination possibilities. Unfortunately, the reactive nature of this alkali metal leads to an exceeded reactivity with the insulating interfaces causing shorts and finally exploding devices. Based on the large numbers of known incidents, the finite lithium deposits and the often high costs one already faces a variety of promising future electrolytes.

One path that is already being pursued includes the exchange of metal salts with ionic liquids which are characterized as non-flammable, non-explosive and above all cheap and high-conductive materials with negligible vapor pressure and high degree of tunability. Nevertheless, their often low viscosities precludes them from many applications. To overcome this drawback, Ohno and coworkers[2–4] demontrated the advantages of a macroscopic architecture of classical polymers with the outstanding features of Ionic Liquids (ILs), resulting in the development of the novel class named Polymeric Ionic Liquids (PILs). After almost 20 years of huge synthetic effort as well as dedicating many studies to the characterization of their physical and mechanical properties, one is still faced with the significant obstacle that PILs have a consequently lower conductivity compared to their small-molecule congeners. Although, a series of papers predict numerous properties, e. g. decoupling of charge transport from segmental dynamics[5–9] or fast proton hopping,[10–12] to be key-features for the design of highly conductive PILs - none of them is able to provide a generally successful strategy, yet.

In this study we address the following major questions by studying the novel PIL, poly[tris(2-(2-methoxyethoxy)ethyl)ammonium acryloxypropyl-sulfonate] (PAAPS):

I How is the molecular dynamics characterized and how does it influence the charge transport mechanism?

II What is the origin of the outstanding high dc-conductivity as well as its novel molecular weight dependent increase?

III Which signature delivers electrode polarization in such complex materials; is it comparable to the case of low molecular weight ILs?

In order to achieve a new level of understanding, Broadband Dielectric Spectroscopy(BDS) covering a wide frequency and temperature range is employed; supplemented by Differential Scanning (DSC) and AC-Chip (ACC) Calorimetry and in addition supported by Infrared (IR) and UV/Vis Spectroscopy.

# Experimental Details

## Measurement Techniques

*Broadband Dielectric Spectroscopy* (BDS) measurements were carried out using a high-resolution $\alpha$-analyzer from NOVOCONTROL Technologies GmbH & Co. KG in a temperature range of $100 - 400$ K and frequency window of $10^{-2} - 10^{7}$ Hz combined with a Quatro temperature controller ensuring absolute thermal stability of $\leq 1$ K. The sample cells for these measurements consist of two blank (mean square roughness $\leq 1$ $\mu$m) plane brass electrodes (lower: spectrometer ground plate ($d = 40$ mm), upper: $d = 10$ mm) which are separated by 2-3 50 $\mu$m glass fiber spacers ordered in parallel. The arrangement of upper electrode, spacers and sample material is afterwords annealed at 150°C inside an oil-free vacuum with $10^{-6}$ mbar for 24h until the upper electrode is placed on top of the liquid like sample droplet. The whole annealing and measurement process is carried out under inert nitrogen or argon atmosphere.

*AC-Chip Calorimetry* (ACC), emplying XEN-39390 chips from Xensors Integration, was conducted using a setup from the group of Prof. Schick at the University in Rostock as described in [13]. The measurements were carried out in a temperature range between -100 $-$ 50°C with rates of 1K/min in a heating and cooling cycle by operating frequencies in a range of $10^{-1} - 10^{4}$Hz. The glass transition temperature at a certain frequency is determined as the midpoint of the step of the magnitude of the measured voltage that is proportional to the real part of the complex heat capacity. The $T_g$s determined in this manner are presented in Table 1.

*Differential Scanning Calorimetry* (DSC) was carried out using Mettler Toledo DSC 1 Star System (temperature rate: 10 K/min). The glass transition temperatures were determined after running two temperature cycles from 80 to -80 °C and reverse. The $T_g$ values were taken as the midpoint of a characteristic heat capacity step upon the dynamic transition from glassy to liquid state.

Table 1: Comparison of the glass transition temperatures $T_g$ measured with DSC, ACC and extrapolated from BDS-data, as well as the dc-conductivity $\sigma_0$ at 300K, and at 50K and 100K, respectively, above $T_g$.

| Mn [g/mol] | $T_g^{DSC}$ [K] | $T_g^{ACC}$ [K] | $T_g^{BDS}$ [K] | log $\sigma_0$ (300K) [S/cm] | log $\sigma_0$ ($T_g$ + 50K) [S/cm] | log $\sigma_0$ ($T_g$ + 100K) [S/cm] |
|---|---|---|---|---|---|---|
| 9700 | 225.2 ± 0.5 | 230 ± 2 | – | −5.5 ± 0.1 | −6.7 ± 0.1 | −4.6 ± 0.1 |
| 44200 | 221.9 ± 0.5 | – | – | −5.4 ± 0.1 | −6.8 ± 0.1 | −4.7 ± 0.1 |
| 51600 | 218.7 ± 0.5 | 224 ± 3 | 225 ± 2 | −5.3 ± 0.1 | −6.8 ± 0.1 | −4.7 ± 0.1 |
| 99500 | 216.4 ± 0.5 | 225 ± 5 | 221 ± 3 | −5.2 ± 0.1 | −6.7 ± 0.1 | −4.7 ± 0.1 |

*Gel permeation chromatography* (GPC) was performed using NOVEMA Max linear XL column with a mixture of 80% of aqueous acetate buffer and 20% of methanol. Conditions: flow rate 1.00 mL/min, PSS standards using RI detector-Optilab-DSP-Interferometric Refractometer (Wyatt-Technology).

*$^1$H Nuclear Magnetic Resonance* (NMR) spectra were performed on a Bruker DPX-400 spectrometer in deuterated solvents using residual solvent as reference.

*Analytical Ultracentrifugation* was employed to determine the partial specific volume of the samples in a density oscillation tube (DMA 5000M, Anton Paar, Graz). The sedimentation experiments were performed on an Optima XLI centrifuge (Beckman Coulter, Palo Alto CA) and Rayleigh interference optics at 25°C and a speed of 60000 rpm. The sedimentation coefficient distributions have been evaluated with the least squares g*(s) evaluation implemented in the software SEDFIT. For equilibrium experiments seven concentrations have been analyzed at different speeds starting from 7500 rpm up to 30000 rpm. Data were evaluated with the program MSTAR.

*UV-VIS Spectroscopy* (UV/Vis) was conducted using the UV-visible spectrophotometer UV1 from Thermo Fisher Scientific in a wavelength range between 200 and 1000 nm. The PAAPS-samples for these measurements were dissolved in highly pure ethanol at a concentration of 4 $\mu$mol/mL and shaken for about 12 hours to achieve a homogeneous solution.

*Geometry Optimization* is performed employing the geometry optimization implemented in the Avogadro program with the Merck molecular force field (MMFF94).[14]

## Synthesis

*Monomer Synthesis* was prepared using a modified procedure as described in [15]. Amberlite resin (IR-120, 1.8 meq/mL) was soaked in a 1.5 fold excess of tris[2-(2-methoxyethoxy)ethyl]amine in water (1:1 v/v) for 24 hours. The acid-base reaction between the acidic resin and the amine resulted in a slight warming of the vessel and ionically attached the amine. The resin was then rinsed with deionized water until a neutral pH was obtained from the rinsing solvent and stored wet in a sealed vessel. An aliquot of wet resin (60 g) was rinsed with acetone to remove water, added to a suspension of potassium 3-sulfopropyl acrylate (12 g, 51.7 mmol) in acetone (200 mL) and shaken using a wrist-action shaker for 5 hours. Ion-exchange was monitored using $^1$H NMR spectroscopy by comparing the integration of the methylene protons on the sulfonate anion ($\delta$ = 4.20, $CH_2CHCOOCH_2$-) to the methylene protons on the ammonium cation ($\delta$ = 3.77, $NCH_2$–). The reaction was deemed complete when the integration ratios of these two signals were 1:3. The solution was then filtered and solvent evaporated *in vacuo* to isolate the monomer (22.3 g, 90%).

*Polymer Synthesis* followed independently of the molecular weight a general procedure. Monomer (14.5 g, 28.0 mmol) was dissolved in DMF (20 mL) in a roundbottom flask followed by the addition of the RAFT agent 2-phenyl-2-propyl benzodithioate (0.026 g, 0.1 mmol) and azobisisobutyronitrile (AIBN, 0.012 g, 0.07 mmol). The solution was then purged with $N_2$ for 20 minutes and heated to 75°C. A pseudo-first order curve was plotted for the polymerization over time and demonstrates the controlled nature of the process (Scheme 1). At 80% conversion, a target molecular weight of 116 000 g/mol could be theoretically achieved. Different molecular weights were thus obtained by halting the reaction (by quickly cooling to room temperature) at different times and analyzed by GPC to measure molecular weights. The polymer solutions were then dialysed (1 kDa cutoff) against deionized water to remove monomer and solvent evaporated *in vacuo* to isolate the pure polymer.

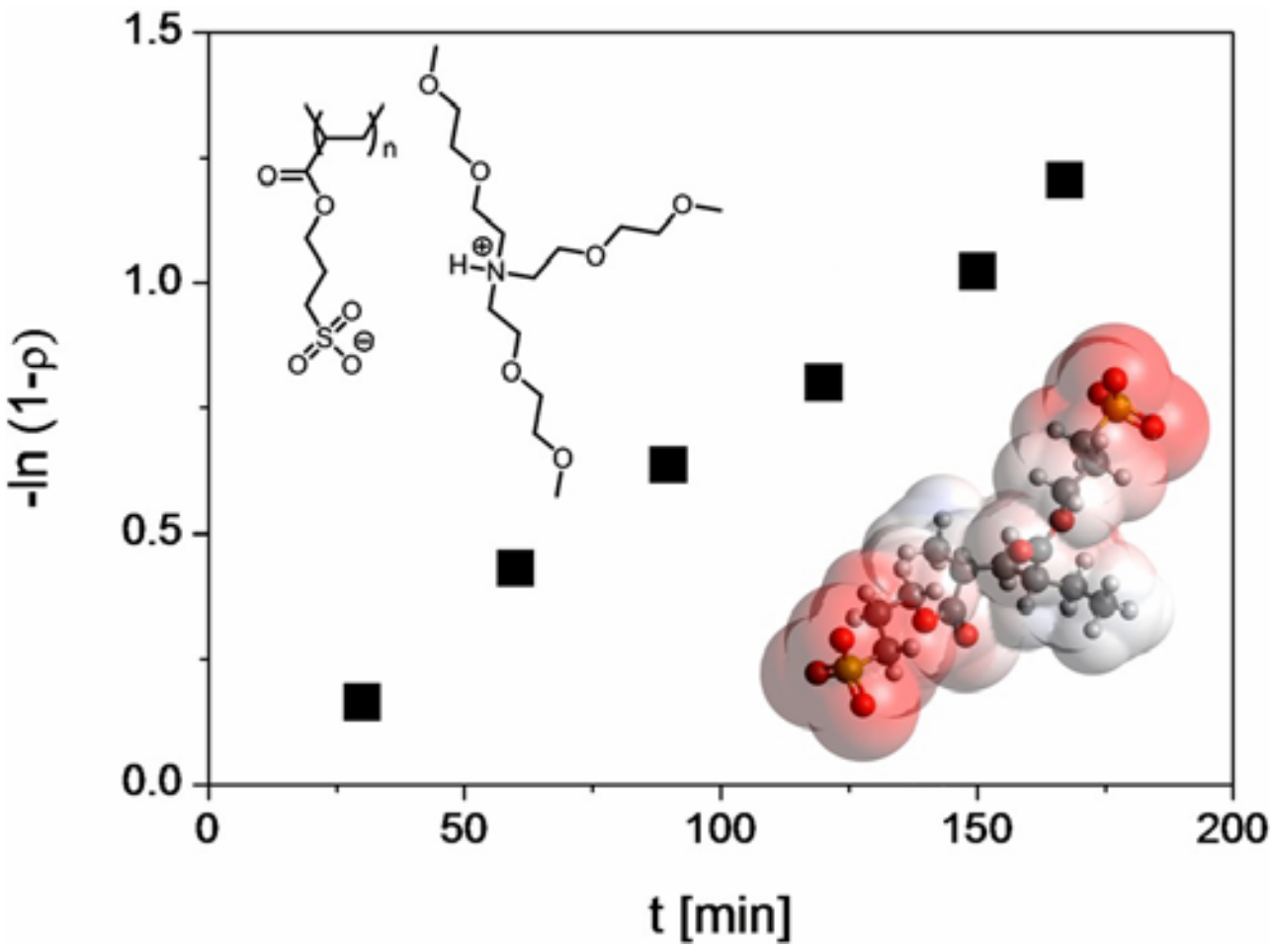

Scheme 1: Demonstration of the controlled RAFT polymerization following a pseudo first-order kinetics; $\rho$ displays the fractional conversion of monomer as determined by $^{1}$H NMR spectroscopy. The upper left inset displays the chemical structure of PAAPS under study, the lower right the van der Waals-surface colored by the electron density of the molecule, representing a dimer.

In this study, four different samples with molecular weights of 9 700 g/mol, 44 200g/mol, 51 600 g/mol and 99 500 g/mol are investigated. Their reaction times were 20, 45, 60, and 170 minutes respectively. The maximum polydispersity index (PDI) values were determined by analytical ultracentrifugation to D=1.43, 1.31, 1.59, and 2.11 respectively (c. f. Supporting Information).

*Synthesis of PAPS*: PAAPS with a molecular weight of 99500 g/mol (5 g) was dissolved in a 1 M solution of aqueous ammonium chloride (100 mL), added to a dialysis tube (1 kDa cutoff), and immersed in a 1 M solution of aqueous ammonium chloride (1 L). After 3 days, the solution was replaced with fresh water multiple times to remove excess salts. The dialysis tube was emptied in to a roundbottom flask and freeze dried, leaving a pink solid identified as the $NH_4^+$-functionalized PAPS (0.6 g, 29%).

# Results and Discussion

The dielectric spectra of the polymeric ionic liquids under study are characterized by a superposition of (i) molecular fluctuations, (ii) charge transport and (iii) electrode polarization (EP). The first, arise from inter- and intramolecular dipole relaxations, and are best observable in $\varepsilon''$ *vs.* fre-quency (Fig. 1) below room temperature. Havriliak-Negami-functions are fitted to this absorption processes, from which the mean relaxation times $\tau_\alpha$ and $\tau_\beta$ are extracted. Charge transport is characterized by the dc-conductivity $\sigma_0$ that shows up as distinct plateau in $\sigma^{/}$ *vs.* frequency or slope in $\varepsilon^{//}$ *vs.* frequency (Fig. 1) as well as the critical frequency $\omega_c$ being interpreted as charge carrier hopping rate which marks the onset to a power law dependence $\sigma^{/} \sim \omega^x$. Electrode polarization caused by the accumulation of ionic charge carriers at the electrode-sample interface contributes as well.

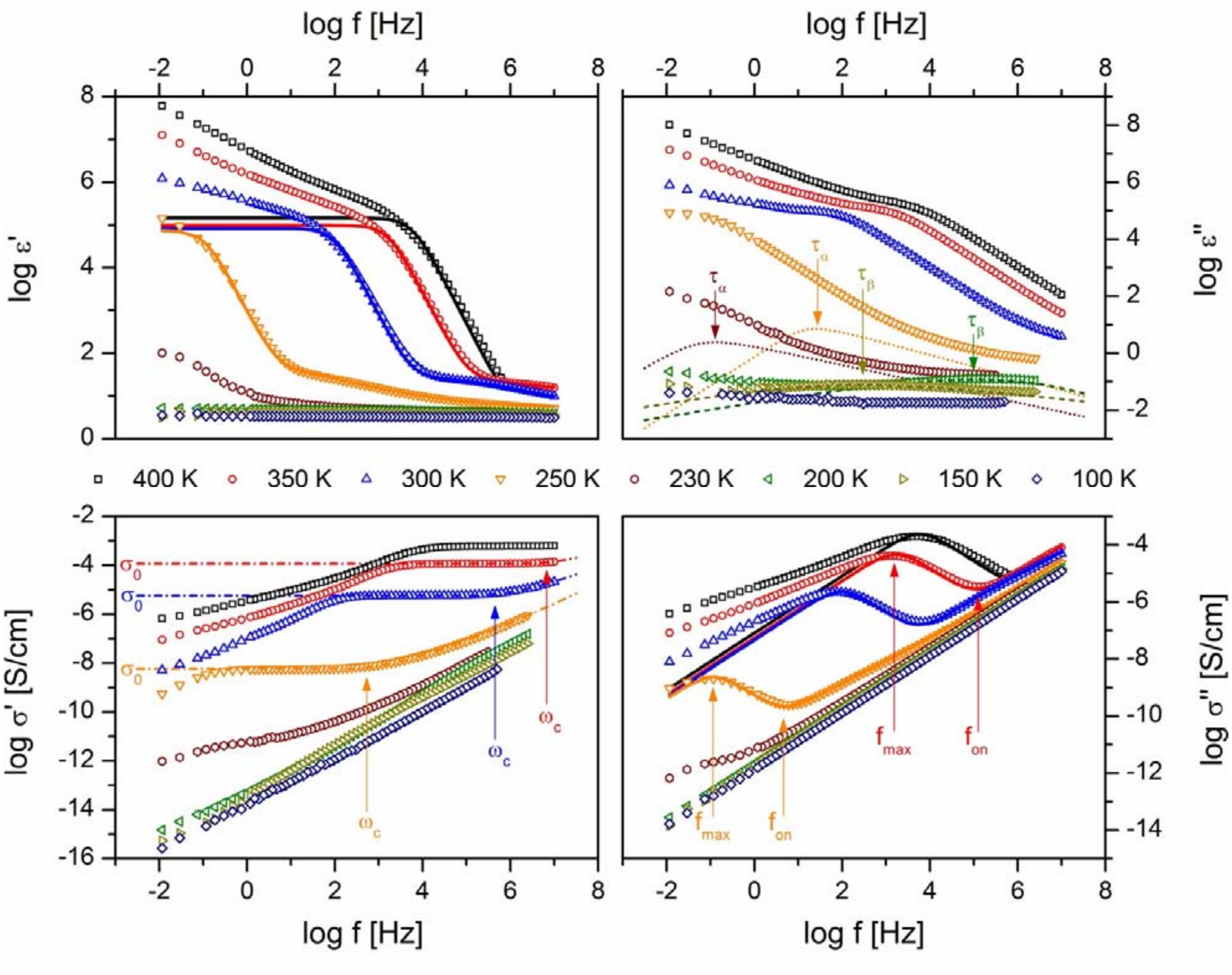

Figure 2: Real and imaginary part of the complex dielectric function $\varepsilon = \varepsilon' - i\varepsilon''$ and complex conductivity $\sigma = \sigma' + i\sigma''$ *vs.* frequency of the PAAPS sample with a molecular weight Mn = 51600/mol at eight different temperatures as indicated. The low frequency regime is dominated by electrode polarization (EP) that is best observable as a steep increase in $\varepsilon'$ *vs.* f or $\sigma''$ *vs.* f. The straight lines indicate a model function for a double layer accumulation at the metal electrode following the model of Serghei et al. with the characteristic frequencies $f_{on}$ and $f_{max}$. $\sigma'$ *vs.* f reveals a distinct plateau representing the dc-conductivity value $\sigma_0$ and the charge carrier hopping rate $\omega_c$ which marks the onset to a power law dependence $\sigma' \sim \omega^x$. This behavior can be fully described by the Dyre-formula (dashed-dotted line). In addition, one observes in $\varepsilon''$ *vs.* f representation two dielectrically active relaxation processes ($\tau_\alpha$ : dotted line, $\tau_\beta$ : dashed line) at lower temperatures. Fits using Havriliak-Negami functions deliver the mean relaxation rates $\tau_\alpha$ and $\tau_\beta$. The logarithm is to base 10; the error bars are smaller then the size of the symbols, unless otherwise indicated.

To access the molecular dynamics of the PILs under study the absorption processes in the imaginary part of the complex dielectric function (Fig. 1) are fitted with generalized relaxation functions according to Havriliak and Negami[16]

$$\varepsilon_{\mathrm{HN}}(\omega) = \varepsilon_\infty + \frac{\Delta\varepsilon}{(1+(i\omega\tau_{\mathrm{HN}})^\beta)^\gamma}, \qquad (1)$$

where $\Delta\varepsilon$ represents the dielectric strength, $\varepsilon_\infty = \lim_{\omega\tau \gg 1} \varepsilon'(\omega)$, $\alpha$ and $\beta$ are shape parameter and $\tau_{\mathrm{HN}}$ the mean relaxation time of all molecular dipoles contributing to a certain relaxation process. For the PAAPS samples under study two dielectrically active relaxation processes are observed and their so determined mean relaxation times $\tau_\alpha$ and $\tau_\beta$ examined in dependence on the inverse temperature 1000/K (Fig. 2). The first dielectrically active relaxation process is partly masked by a conductivity contribution (c. f. Supporting Information) that restricts its analysis to a very narrow temperature range for two samples (Fig. 2). In order to overcome this limitation ac-chip calorimetry (ACC), which is not effected by any conductivity signal and able to measure the evolution of the dynamic the glass transition was employed in a range between $10^0$ and $10^4$ Hz. Additionally, DSC measurements were conducted delivering the calorimetric glass transition temperature $T_g^{DSC}$ at the conventional relaxation time of 100 s. Both, the structural relaxation rates $\tau_\alpha$, either deduced from BDS or ACC coincide, and the glass transition temperatures extrapolated by the latter are in agreement with those from DSC; thermal deviations are presumably caused by device-related temperature rates (DSC: 10 K/min, ACC: 1 K/min, BDS: in temporary equilibrium).

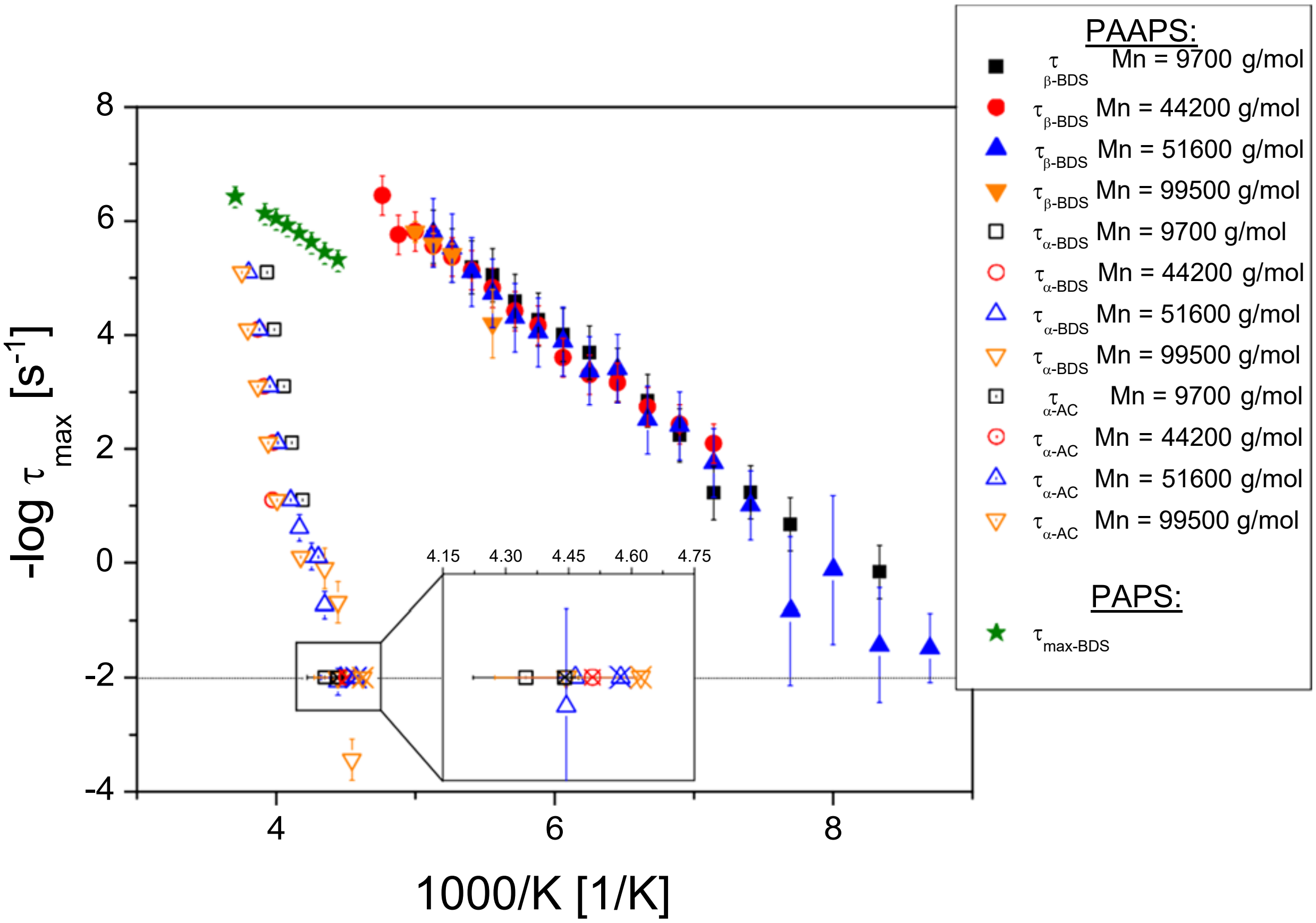

Figure 3: Activation plot of both, structural relaxation rates ($\tau_\alpha$: open symbols, $\tau_\beta$: filled symbols) for all four PAAPS-samples measured by BDS, as well as the dynamic glass transitions at various frequencies derived by ACC (open symbols with dots) and the calorimetric glass transition temperatures obtained by DSC (open symbols with cross). In addition, the structural relaxation rate ($\tau_{\max}$) of the modified chemical structure PAPS (stars) is presented. The logarithm is to base 10; the error bars are smaller than the size of the symbols, unless otherwise indicated.

The thermal activation of $\tau_\alpha$ follows a Vogel-Fulcher-Tammann temperature dependence leading to the assignment of this process to the dynamic glass transition; in accordance with the literature[6,9,17–19] tentatively originating from the relaxation of dipoles between mobile charge carriers. The secondary relaxation with relaxation time $\tau_\beta$ is well separated and - within measurement accuracy - identical for all PAAPS samples. It appears at temperatures below 225K and follows an Arrhenius-law with a constant activation energy $E_\beta = 6 \pm 0.5 \cdot 10^{-23}$ J $\hat{=}$ 0.37 ± 0.04 meV. In order to determine whether this process originates from fluctuations of the carboxyl ester group within the sidechain or from the oxygen atoms within the cation, the sample was modified to PAPS (where the cation is exchanged with an ammonium cation). By analyzing the dielectric spectra a comparable relaxation to $\tau_\beta$ is found which allows determing the molecular assignment of this process to fluctuations of the carboxyl acid ester group.

The charge transport is characterized by the dc-conductivity $\sigma_0$ and charge carrier hopping rate $\omega_c = \frac{1}{\tau_e}$, which are both analyzed by the Dyre-formula[20]

$$\sigma(\omega) = \sigma_0 \frac{i\omega\tau_e}{\ln(1 + i\omega\tau_e)} \quad (2)$$

The relation between this microscopic hopping process occurring on a sub-nm length scale and the macroscopic conductivity measured over the whole sample cell is well known for low molecular weight ILs.[21–26] It is remarkable that the Barton-Namikawa-Nakajima (BNN)-relation $\sigma_0 \sim \omega_c$ is fulfilled over many orders of magnitude (Fig. 3). Moreover, $\sigma_0$ as well as $\omega_c$ correlate strongly with the structural relaxation rate $\tau_\alpha$ (Fig. 3 - inset) conclusively assigning this molecular process to a relaxation of ionic charge carriers responsible for the charge transport and likewise origin of the dynamic glass transition. Thus, one is able to determine the charge transport mechanism as a dynamic glass transition assisted hopping process within a random potential landscape. Beside the fact that $\tau_\alpha \sim \omega_c$ holds for all PAAPS samples, it is worth mentioning that the values for $\omega_c$ increase by about 2 orders of magnitude by increasing the molecular weight from 9700 to 99500 g/mol at the same state of structural dynamics $\tau_\alpha$ ; implying a sterically supported hopping conduction.

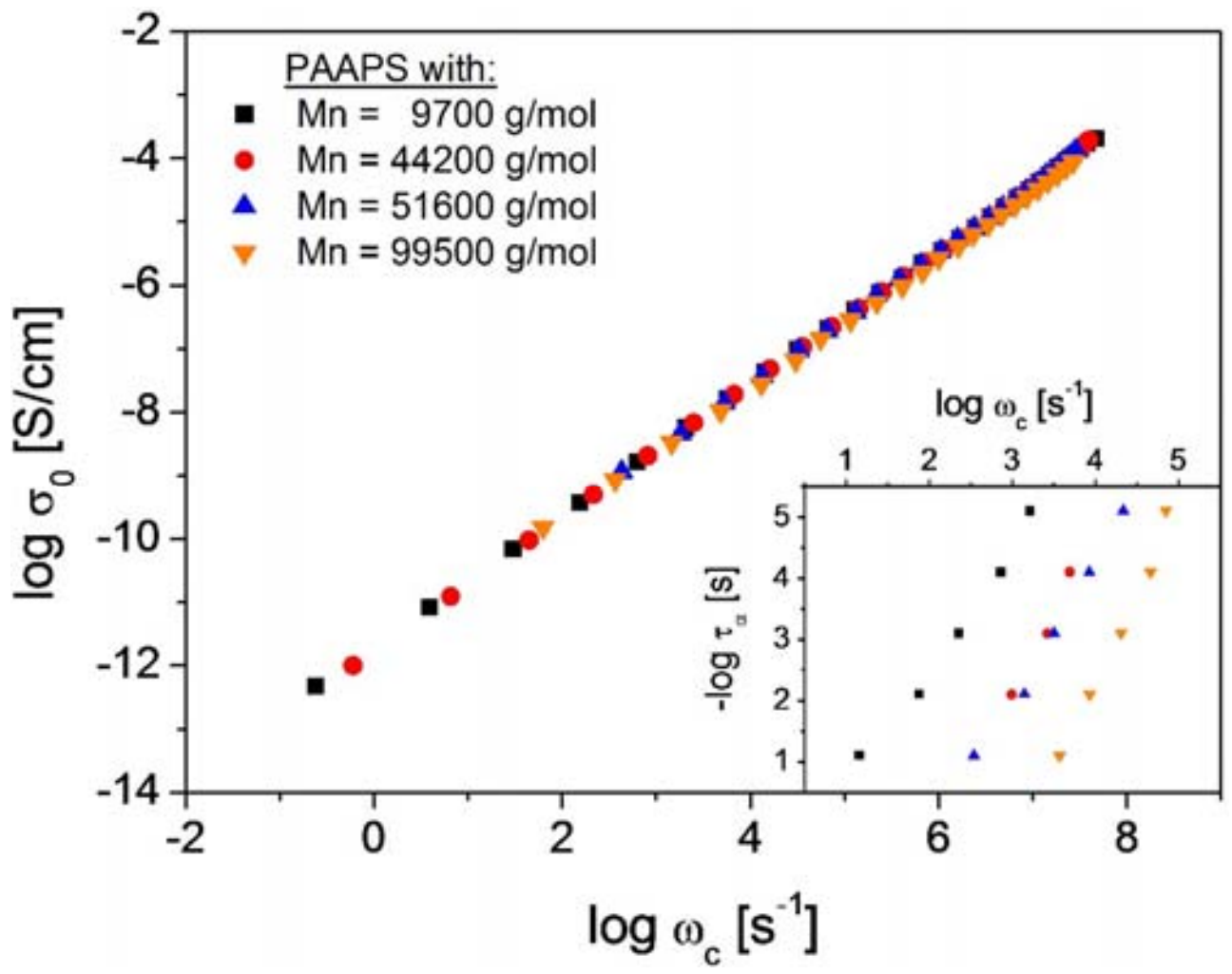

Figure 4: Barton-Namikawa-Nakajima relation $\sigma_0 \sim \omega_c$ between the dc-conductivity $\sigma_0$ and the charge carrier hopping rate $\omega_c$ for all four PAAPS samples as indicated. The inset shows the correlation between the latter and the structural relaxation rate $\tau_\alpha$ of the transient dipoles. The logarithm is to base 10; the error bars are smaller than the size of the symbols, unless otherwise indicated.

In order to unravel this steric support, UV/Vis-spectroscopy is employed (Fig. 5). For the PAAPS sample with the lowest molecular weight, the UV/Vis absorption spectrum shows a broad peak at around 300 nm, whereas for the high molecular weight PIL one obtains a distribution of superimposing absorption processes. In case the electronic structure of one polymer repeating unit was not affected by the adjacent units, the UV/VIS absorption appears identical for all samples regardless of their molecular weight. Because the absorption pattern of the high MW polymer is broader and shows a sophisticated texture it is evident that this PIL allows for more UV/Vis-active transitions in a wider wavelength range. These transitions inevitably arise from interaction of the electronic structure between neighboring side chains, at which the highest charge density and the HOMO-niveaus are located. Consequently the interaction between the electron states of neighboring side chains favors charge transport in the direction of the polymer main chain. In addition, vibrational spectroscopy reveals an enhanced degree of hydrogen bonding in the high polymeric system compared to the one with a molecular weight of 9700 g/mol (c.f. Supporting Information). As discussed in the literature,[27–32] it is likely that hydrogen bonding supports charge transport.

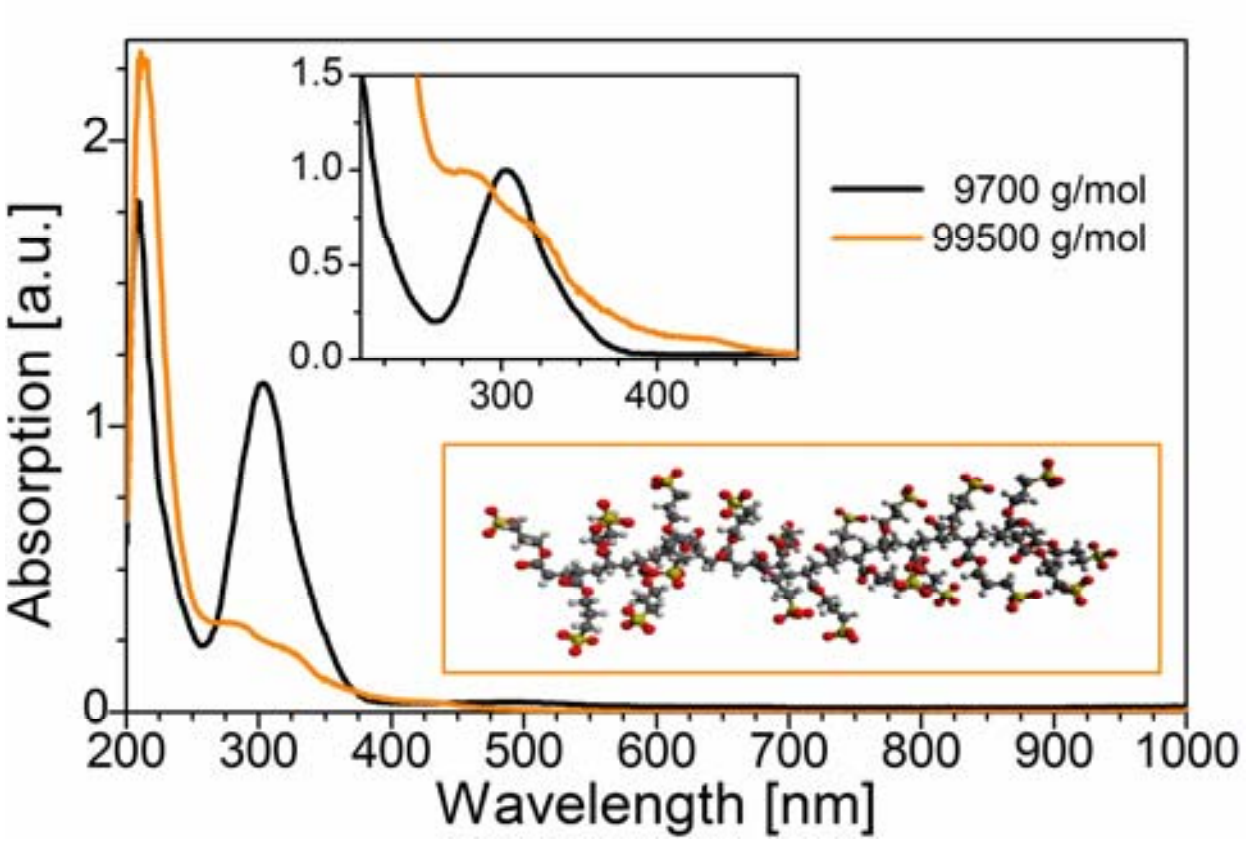

Figure 5: UV/Vis-absorption spectra for the PAAPS samples with the lowest (9700 g/mol) and highest (99500 g/mol) molecular weight as indicated. The upper central inset shows pronounced differences between the two samples, in which most evolved peaks are rescaled to 1; the lower central inset display the elongated mesoscopic structure for an isolated molecule with the molecular weight of 9700 g/mol simulated with the MMFF94 force field.

Beside molecular relaxation processes and charge transport, electrode polarization plays a prominent role on the low frequency side; it is characterized as steep increase at high temperatures for $\varepsilon'$ *vs.* frequency and $\sigma''$ *vs.* frequency (Figs. 1, 5). Compared to low molecular weight, e. g. imidazolium or pyrrolidinium based ILs,[33–35] one observes consistently an onset frequency $f_{on}$ at which $\varepsilon'$ or $\sigma''$ drastically increases, but no longer a maximum frequency $f_{max}$ (adopted nomenclature - ref.[33]), at which the evolution of this process is saturated. In fact, assuming a double layer thickness of 3 nm and a mobility decrease of $10^{-4}$ compared to bulk, the dielectric signature is described by the model of Serghei et. al[33,35] up to $f_{max}$ (model fits in Fig. 5). Because of the heterogeneous nature of the PIL deviations on the low frequency side are observed. Since an applied voltage of 0.1 V is used for the measurements (which translates to 200 V/cm, given the inter-electrode separation of 50$\mu$m), it follows that these observations cannot be explained by invoking nonlinear effects (Fig. 5 - lower right insets).

Furthermore, temperature dependent deviations of the EP are evident which do not appear for low molecular weight ILs. The EP signature grows stronger with temperature (Fig. 5 - upper right insets) suggesting the development of more than one interfacial double-layer due to the thermal activation of the ions and a weaker shielding of the bulk material from the metal electrodes. The latter is deduced from the about 1 - 1.5 decades smaller step in $\varepsilon^{/}$ compared to low molecular weight ILs.[24,33,35,36]

For all samples studied in this work, one observes - beside a similar behavior above frequencies of $f_{max}$ - two systematic molecular-dependent variations: (I) the characteristic frequencies ($f_{max}$, $f_{on}$) are shifted to approximately one decade higher values at low temperatures, e. g. 250K, for the high molecular weight PIL compared to the one with 9700 g/mol. This is determined by the approximately one decade higher dc-conductivity $\sigma_0$ (Fig. 6) (which is discussed later) of the PIL according to the following relation:

$$\tau_{EP} = \frac{\varepsilon_s \varepsilon_0}{\sigma_0} \frac{D}{2L_D}, \qquad (3)$$

whereat $\tau_{EP} \sim f_{max} \sim f_{on}$, $\varepsilon_s$ presents the saturation value of the EP in $\varepsilon^{/}$, $\varepsilon_0$ is the dielectric constant, D the sample thickness, and $L_D$ the Deby-length.[16] (II) with rising molecular weight the deviation from the Serghei-model is more pronounced (Fig. 2 - upper right insets). Although the relation $E_{coulomb} >> E_{kT}$ holds true for Coulomb interactions between ions and metal electrodes within a distance up to ~ 5-10 nm, the PILs - especially their polyanions - can no longer be considered as hard spheres, for which it is easier by their shape to reaccumulate with the alternating field. In contrast, it is much more likely that PILs develop the first double-layer with more defects compared to ILs due to sterical hindrances which is confirmed by the about one (and a half) decade(s) less pronounced step in $\varepsilon^{/}$ and results in a weaker shielding of the bulk material from the electrode. Consequently, additional - even more heterogeneous - double layers are developed causing deviations from the model of Serghei et al. in the dielectric spectra.

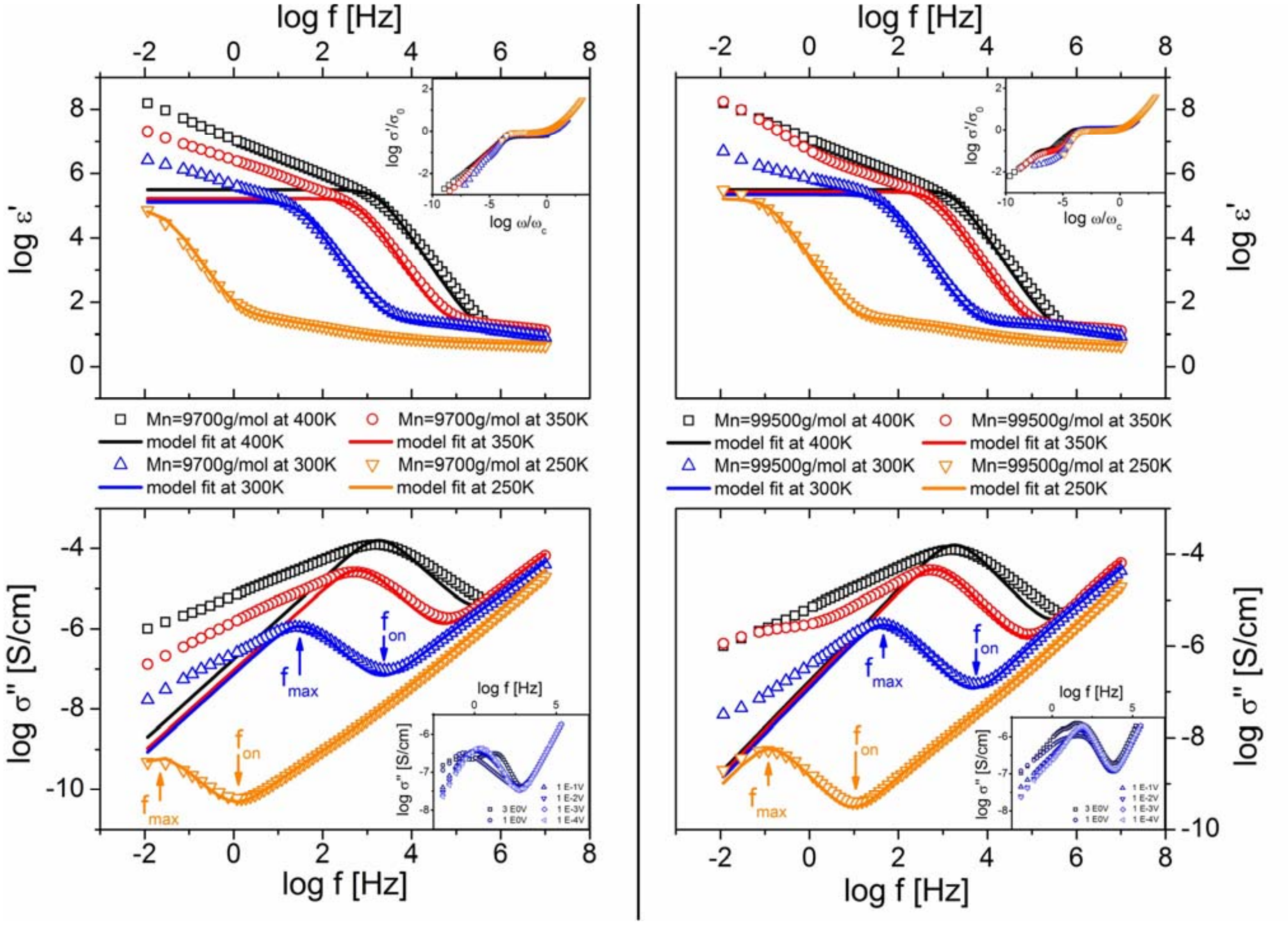

Figure 6: The signature of electrode polarization (EP) as displayed in terms of $\varepsilon^{/}$ *vs.* f and $\sigma^{//}$ *vs.* f for the PIL with the lowest (9700 g/mol - left part) and highest molecular (99490 g/mol - right part) weight at 4 given temperatures as indicated. The straight lines display model fits by the approach of Serghei et al.[33,35] assuming an interfacial region with a thickness of 3nm at the metal electrode whereat the ions mobility is reduced by a factor of $3\times10^4$ compared to bulk. The upper right insets present molecular weight dependent differences of the EP signature in terms of $\sigma^{/}$ *vs.* $\omega$ rescaled to the dc-conductivity $\sigma_0$ and charge carrier hopping rate $\omega_c$. The lower right inset displays the subtle dependence of EP to the applied voltage; showing non-linear effects for field strengths above 200V/cm (by given electrode distances of 50$\mu$m).

Although, the dielectric data of the PAAPS samples under study show a signature comparable to many other PILs, these materials exhibit the highest dc-conductivity $\sigma_0$ values at temperatures below 100 °C of all pure PILs investigated so far. Thus, these results are important from various viewpoints: On the one hand they outline essential achievement towards improved and novel applications, as medium in batteries, solar cells, gas separator membranes or supercapacitors. On the other hand the increasing dc-conductivity with rising degree of polymerization represents an exceptional and contrary result compared to any other samples in this class. Nevertheless, these findings can be potentially adopted in other fields of polymer research leading to more sophisticated materials. A conclusive explanation of these effects will not be possible before atomistic simulations of such complex systems in the condensed state are technically feasible.

In the following the authors will present further facts that, by their current state of knowledge, substantiate this novel behavior. When compared with all chemical structures of representative PILs (Fig. 6 and Supporting Information) it is evident that PAAPS is the only sample whose polyions charge is located at the outermost end of the polymer side chains. Furthermore the charge density at those moieties is significantly greater than for comparable PILs. Additionally, PAAPS is the only sample with a flexible and comparably big counterion. The authors' view is that the polymer mainchain is presumably elongated and, without additional alkyl units at the end of each side chain as in the case of other PILs (Fig. 6), occurs as a rod whose surface is almost homogeneously charged. Additionally, the increased size and the flexibility of the cation supports a comparably weak ion bond as well as a high mobility. The last point is affirmed by the conductivity of PAPS (by exchanging the cation in PAAPS to $NH_4^+$) that is generally magnitudes of order lower. Regarding the increase in $\sigma_0$ with molecular weight it is nearby that the charges of the polymeric mainchain are more densely packed leading to shorter hopping distances, and hence faster hopping rates. Beside that, it can be assumed that the tentatively elongated backbones act as conductivity channels or paths.

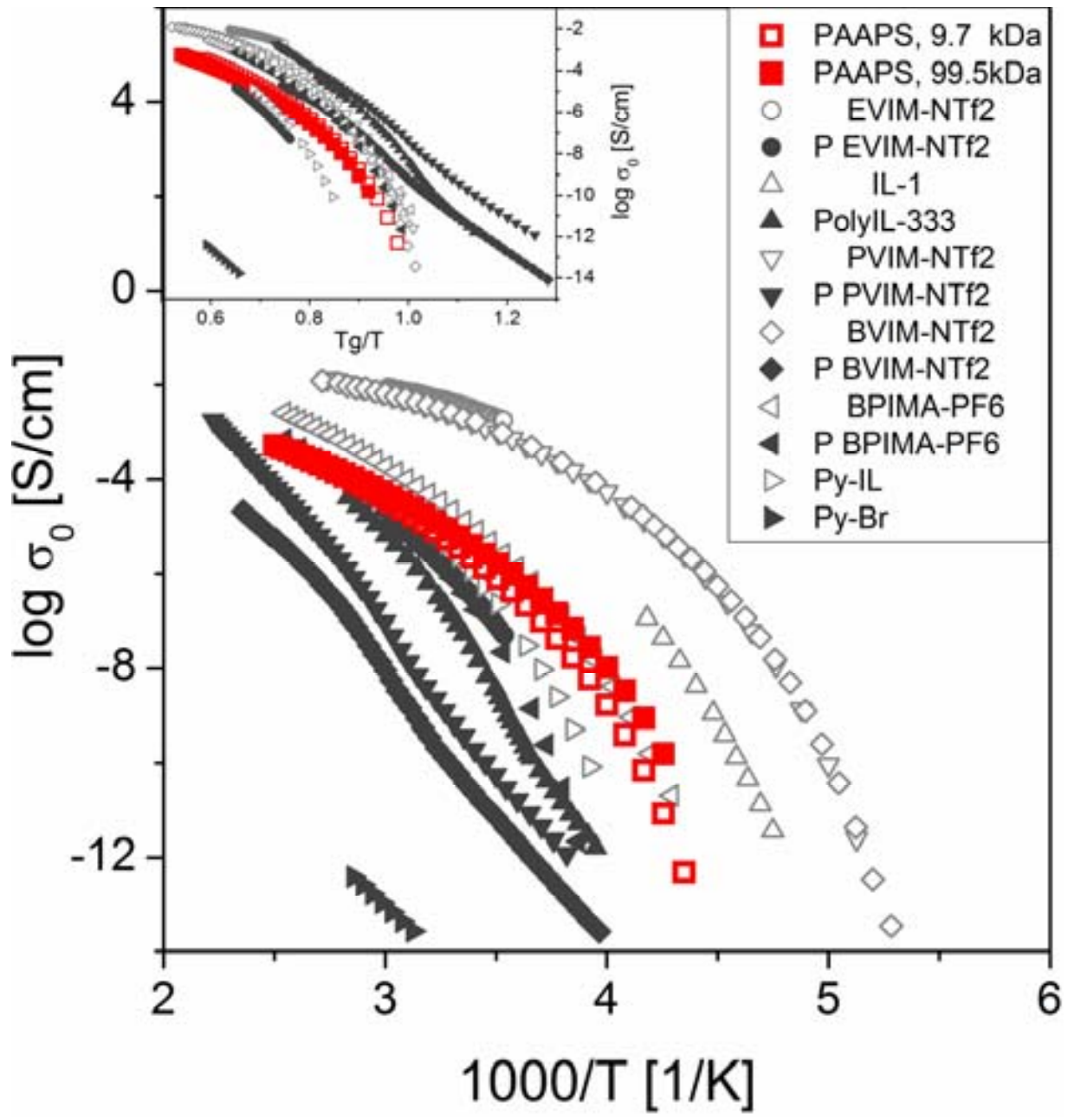

Figure 7: Comparative presentation of the dc-conductivity $\sigma_0$ *vs.* inverse temperature for the PILs with the lowest and highest MWs discussed in this study, as well as six other PILs and corresponding ILs analyzed over wide temperature and frequency range taken from the literature.[4,6–8,17,37] The low molecular weight ILs are pictured as open symbols, the corresponding PILs as filled ones. The inset displays the same data rescaled to the calorimetric glass transition temperature $T_g$. The logarithm is to base 10; the error bars are smaller than the size of the symbols, unless otherwise indicated.

# Conclusion

Despite the class of polymeric ionic liquids (PILs) was created by the pioneering work of Ohno in 1998 to become a promising candidate for many electrochemical application areas their conductivity decline of typically more than 3 decades (at relevant temperatures) from IL to PIL excludes them from use e. g. as battery electrolyte; a comparable dc-conductivity value between IL and PIL has not yet been observed.

In this paper, we strategically employed Broadband Dielectric Spectroscopy (BDS), Differential Scanning (DSC) and AC-Chip Calorimetry (ACC) as well as UV-Vis-spectroscopy to investigate the glassy dynamics, the underlying charge transport mechanism and the interfacial effects between bulk material and metal electrodes while varying the molecular weights of the PAAPS samples un-der study from an oligomer (9700 g/mol) to a true PIL with almost 200 segmental repeating units (99500 g/mol). On the basis of the experimental data one can draw the following conclusions:

(I) Two well separated dielectrically active relaxation processes are observed by BDS. By additionally employing DSC and ACC one ($\tau_\alpha$ ) that follows a VFT-law is assigned to the dynamic glass transition, while the other ($\tau_\beta$ ), whose thermal activation is Arrhenius-like, originates from fluctuations of the carbon acid ester group attached to the polymer mainchain.

(II) The Barton-Namikawa-Nakajima (BNN)-relation is strongly fulfilled over up to 10 decades for all four samples under study. Moreover, one observes a linear correlation between $\tau_\alpha$ and the charge carrier hopping rate $\omega_c$ over 4 orders of magnitude that leads to the identification of the charge transport mechanism as a dynamic glass transition assisted hopping process within a random potential landscape. Though, it is notable that comparing the low MW PAAPS with the one with 99500 g/mol one observes a hundred times higher hopping rate for the latter at the same temperatures which indicates an eased charge transport along the chain contour.

(III) The electrode polarization (EP) reveals the mesoscopic complexity of this class of materials. While the model of Serghei et. al (with an assumed double layer thickness of 3 nm and a mobility decrease of 4 decades) describes the signature well for the high and intermediate frequency regime, pronounced deviations on the low frequency side that increase with increasing temperature and molecular weight are present. Furthermore, one observes a shift to about ten times higher characteristic frequencies ($f_{on}$, $f_{max}$) for the polymeric system with higher molecular weight.

(IV) The PAAPS samples under study show the highest dc-conductivity $\sigma_0$ value for common PILs (considering temperature- and frequency-dependence studies) below temperatures of 100 °C. Furthermore, for the first time ever we observed, in contrast to the expected $\sigma_0$ drop-down of typically 3 to 6 decades, a conductivity increase with the degree of polymerization (i. e. molecular weight). This behavior indicates a sterical support along the presumably elongated polymeric mainchain that can be documented by changes in the electronic structure observed by UV/Vis-spectroscopy, a more developed H-bonding network examined by IR and basic computer simulations.

One may raise the question which chemical design induces highly conductive polymeric Ionic

Liquid. Comparing the PILs under study with them from the literature one predicts: (i) low glass transition temperatures of the IL and polymeric matrix, (ii) a high number density of charge carriers, (iii) ions attached to the polymeric backbone are exposed at the outermost end of a short flexible linker and (iv) high counterions radii shielding its charge, resp. reducing the potential wells depth.

# Acknowledgment

FF and FK acknowledge the financial support from the Deutsche Forschungsgesellschaft under the DFG-project "Neue Polymermaterialien auf der Basis von funktionalisierten ionischen Flüssigkeiten für Anwendungen in Membranen 'Erkenntnistransfer-Projekt' " (KR 1138/24-1); FK and MA within the SFB/TRR 102, Project B8: "Polymers under multiple constraints: restricted and controlled molecular order and mobility", too. R.G. and J. Y. were supported by the Max Planck Society in Germany and the European Research Council Starting Grant (639720-NAPOLI). The authors gratefully acknowledge Heiko Huth and Prof. Christoph Schick from the University in Rostock for their extensive support regarding the ACC measurement set-up and data analysis; as well as Katrin Steinke and Prof. Jörg Matysik from the Leipzig University for conducting the UV-VIS measurements; and Emmanuel Urandu Mapesa for his language assistance.

# Molecular dynamics and charge transport in highly conductive Polymeric Ionic Liquids

Falk Frenzel,*,† Ryan Guterman,‡ A. Markus Anton,† Jiayin Yuan,‡ and Friedrich Kremer†

*Leipzig University, Institute of Experimental Physics I, Linnéstrasse 5, 04103 Leipzig, Germany, and Max Planck Institute of Colloids and Interfaces, Department of Colloid Chemistry, Am Mühlenberg 1 OT Golm, 14424 Potsdam, Germany*

E-mail: falk.frenzel@physik.uni-leipzig.de

## Supplementary Information

### Experimental

*Fourier-transform infrared* (FTIR) spectroscopy measurements are accomplished in transmission mode on a Bio-Rad FTS 6000 spectrometer equipped with a UMA-500 microscope. The liquid samples (9700 g/mol and 99500 g/mol) are squeezed between two $BaF_2$ IR windows, while the intensity is recorded by means of a mercury-cadmium-telluride (MCT) detector (Kolmar Technologies) with a frequency resolution of $2 cm^{-1}$.

*To whom correspondence should be addressed

†Leipzig University, Institute of Experimental Physics I, Linnéstrasse 5, 04103 Leipzig, Germany

‡Max Planck Institute of Colloids and Interfaces, Department of Colloid Chemistry, Am Mühlenberg 1 OT Golm, 14424 Potsdam, Germany

## Figures

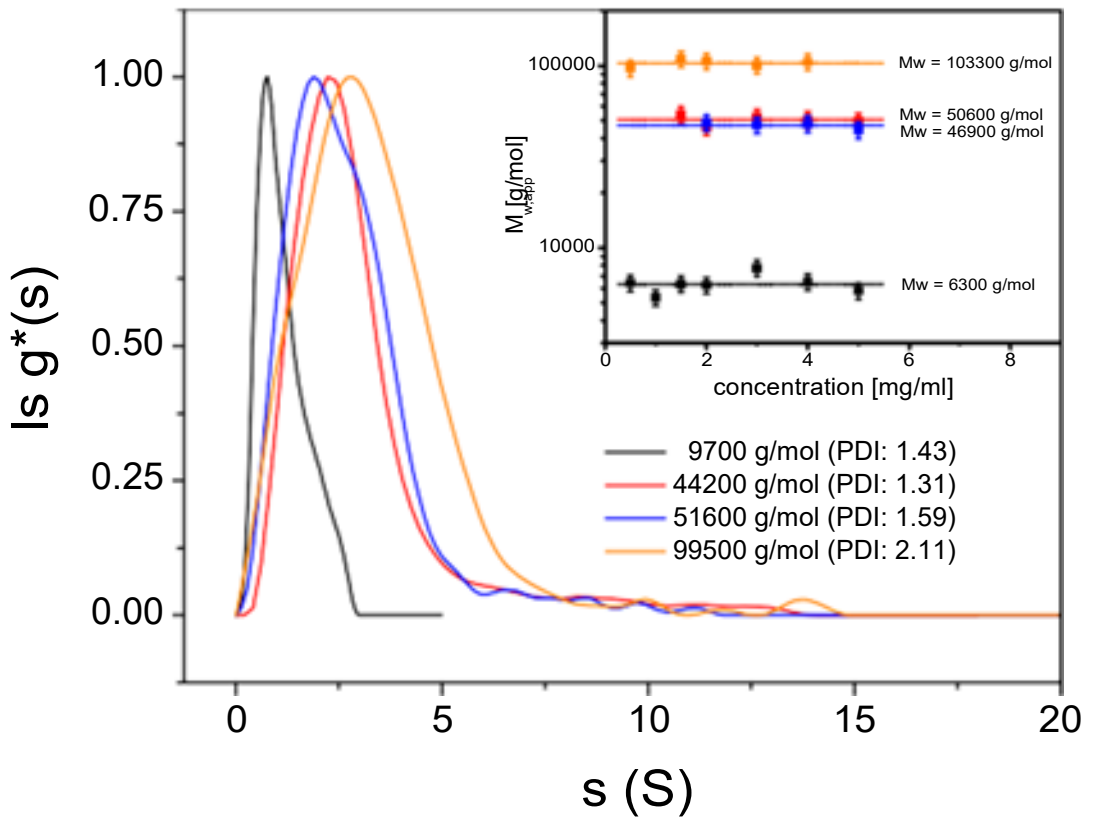

Figure 1: Sedimentation coefficient distribution of the four PAAPS samples with different molecular weights determined by GPC is used to derive the PDI values as indicated. The inset shows the complementary determination of the molecular weights using ultracentrifugation.

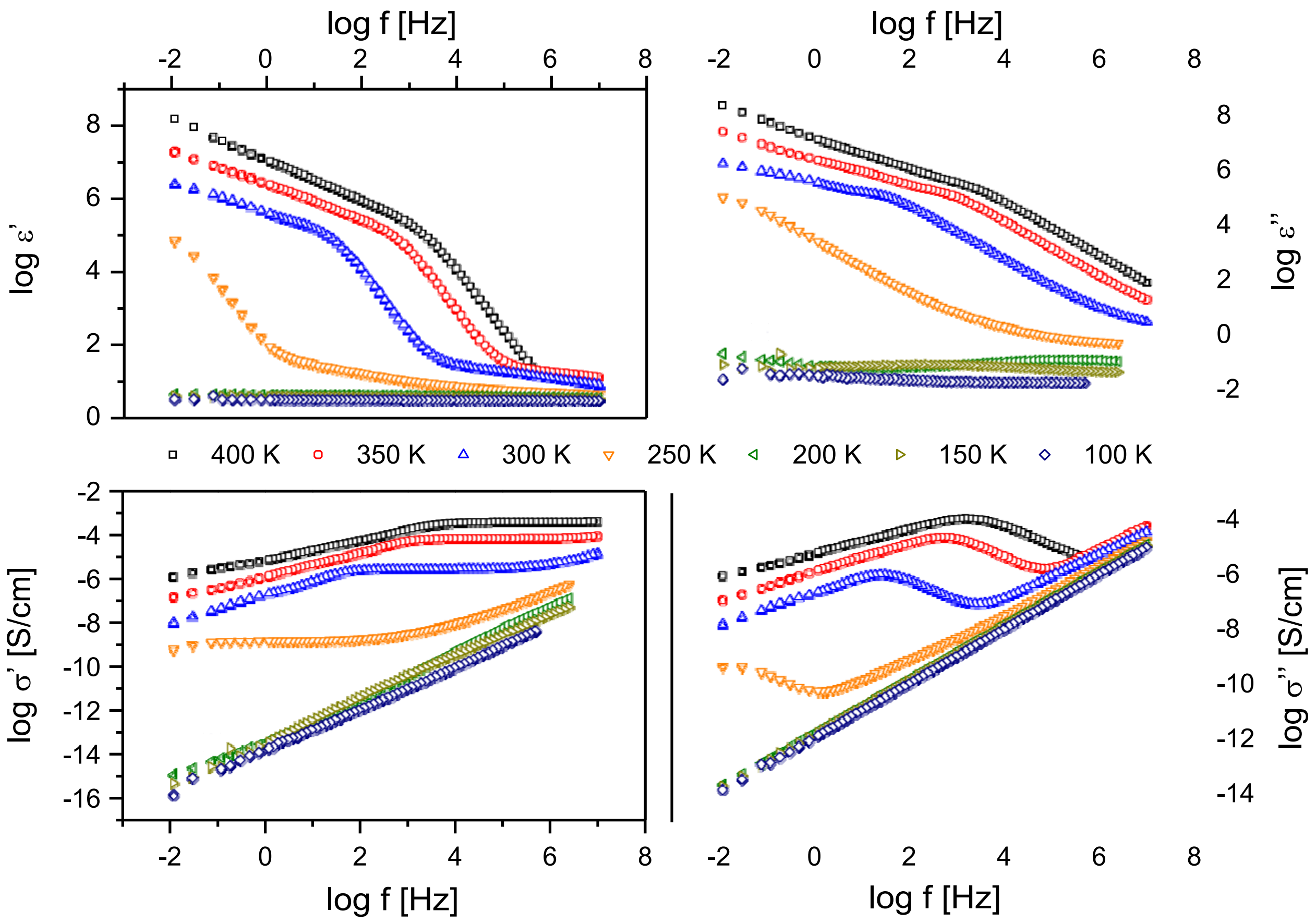

Figure 2: Real and imaginary part of the complex dielectric function $\varepsilon = \varepsilon^{I} - i\varepsilon^{II}$ and complex conductivity $\sigma = \sigma^{I} + i\sigma^{II}$ *vs.* frequency of the PAAPS sample with a molecular weight Mn = 9700g/mol at seven different temperatures as indicated. The logarithm is to base 10; the error bars are smaller then the size of the symbols, unless otherwise indicated.

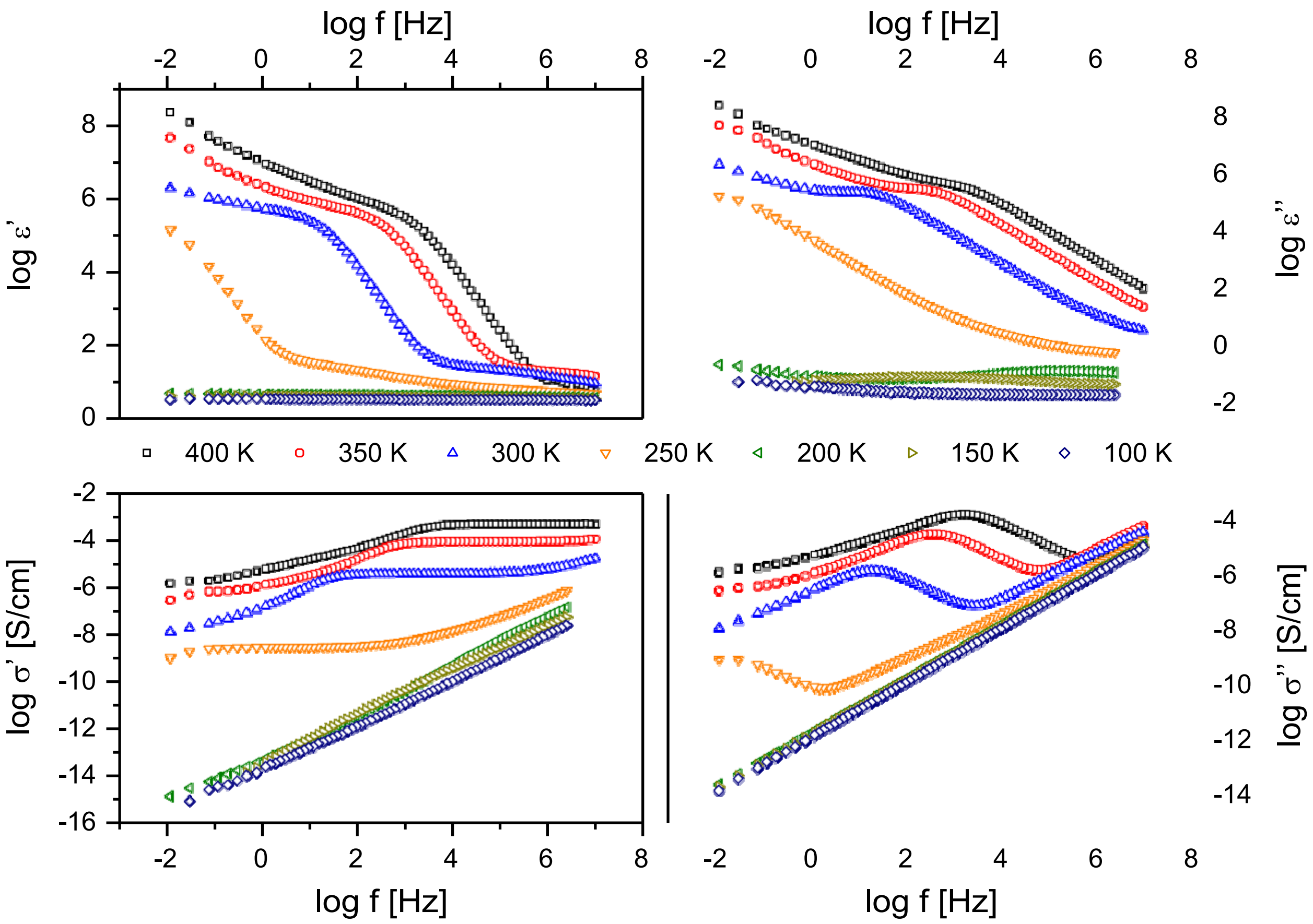

Figure 3: Real and imaginary part of the complex dielectric function $\varepsilon = \varepsilon^{I} - i\varepsilon^{II}$ and complex conductivity $\sigma = \sigma^{I} + i\sigma^{II}$ *vs.* frequency of the PAAPS sample with a molecular weight Mn = 44200g/mol at seven different temperatures as indicated. The logarithm is to base 10; the error bars are smaller then the size of the symbols, unless otherwise indicated.

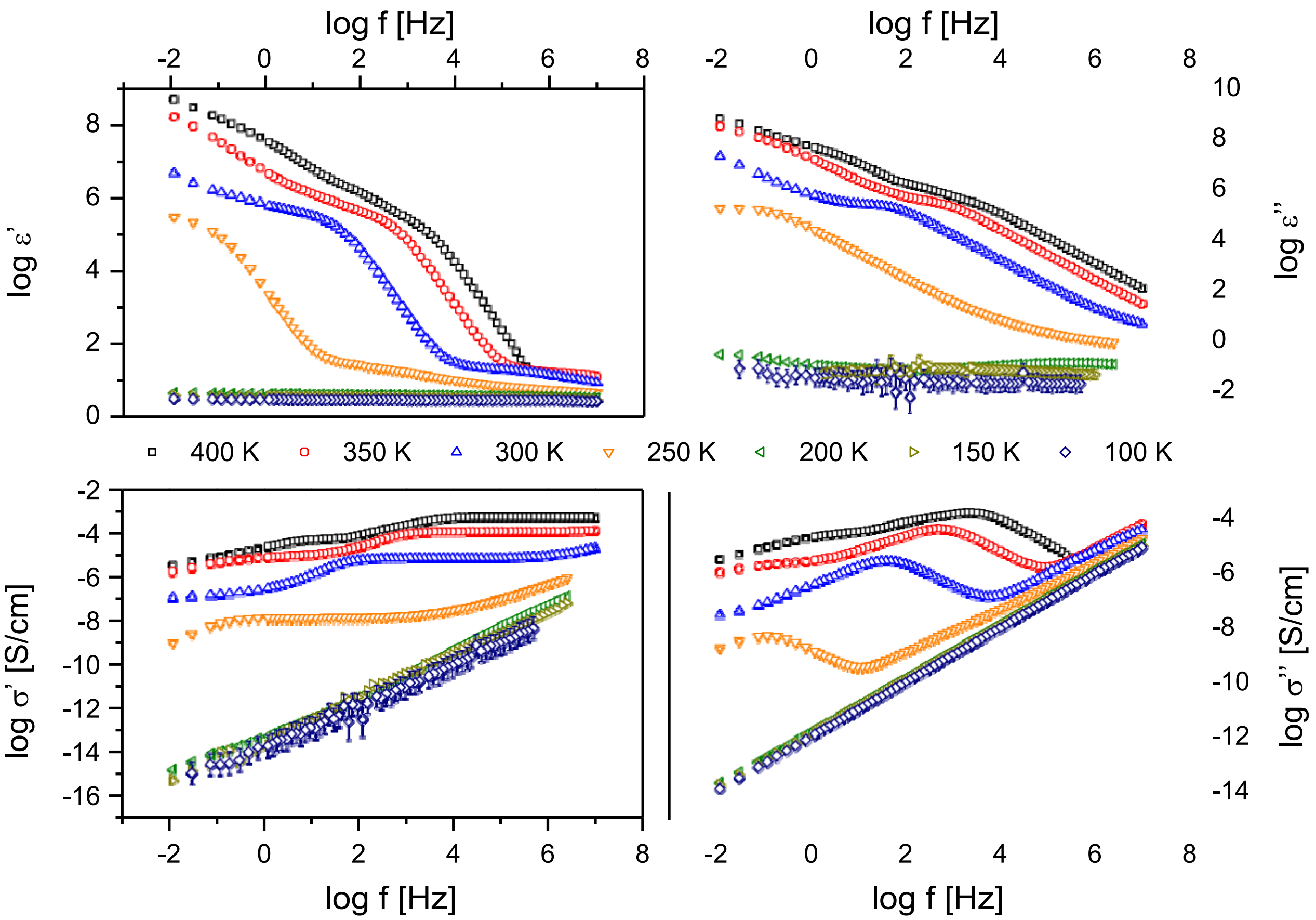

Figure 4: Real and imaginary part of the complex dielectric function $\varepsilon = \varepsilon' - i\varepsilon''$ and complex conductivity $\sigma = \sigma' + i\sigma''$ *vs.* frequency of the PAAPS sample with a molecular weight Mn = 99500g/mol at seven different temperatures as indicated. The logarithm is to base 10; the error bars are smaller then the size of the symbols, unless otherwise indicated.

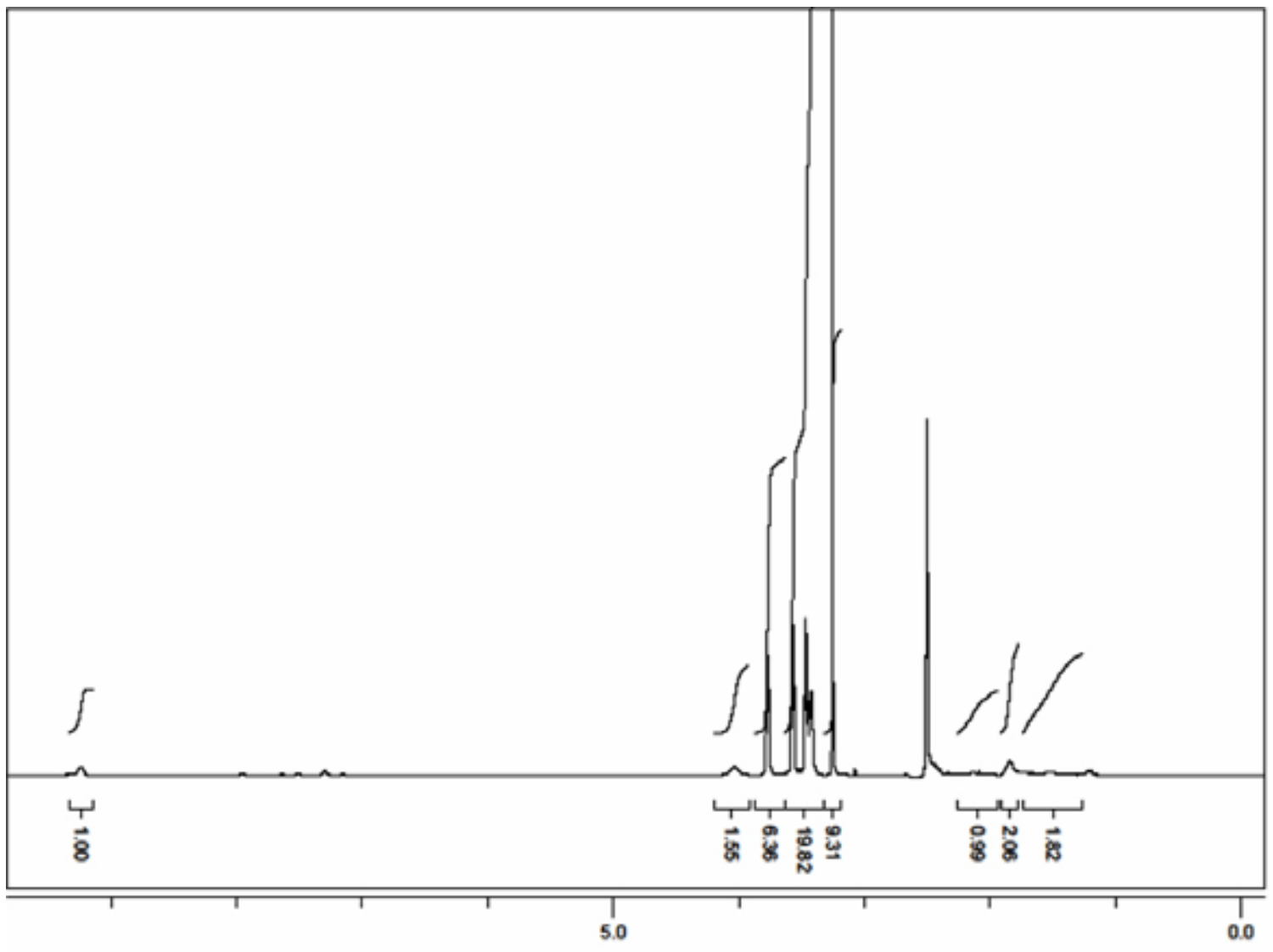

Figure 5: $^1$H NMR spectra of PAAPS with a molecular weight of 9700 g/mol.

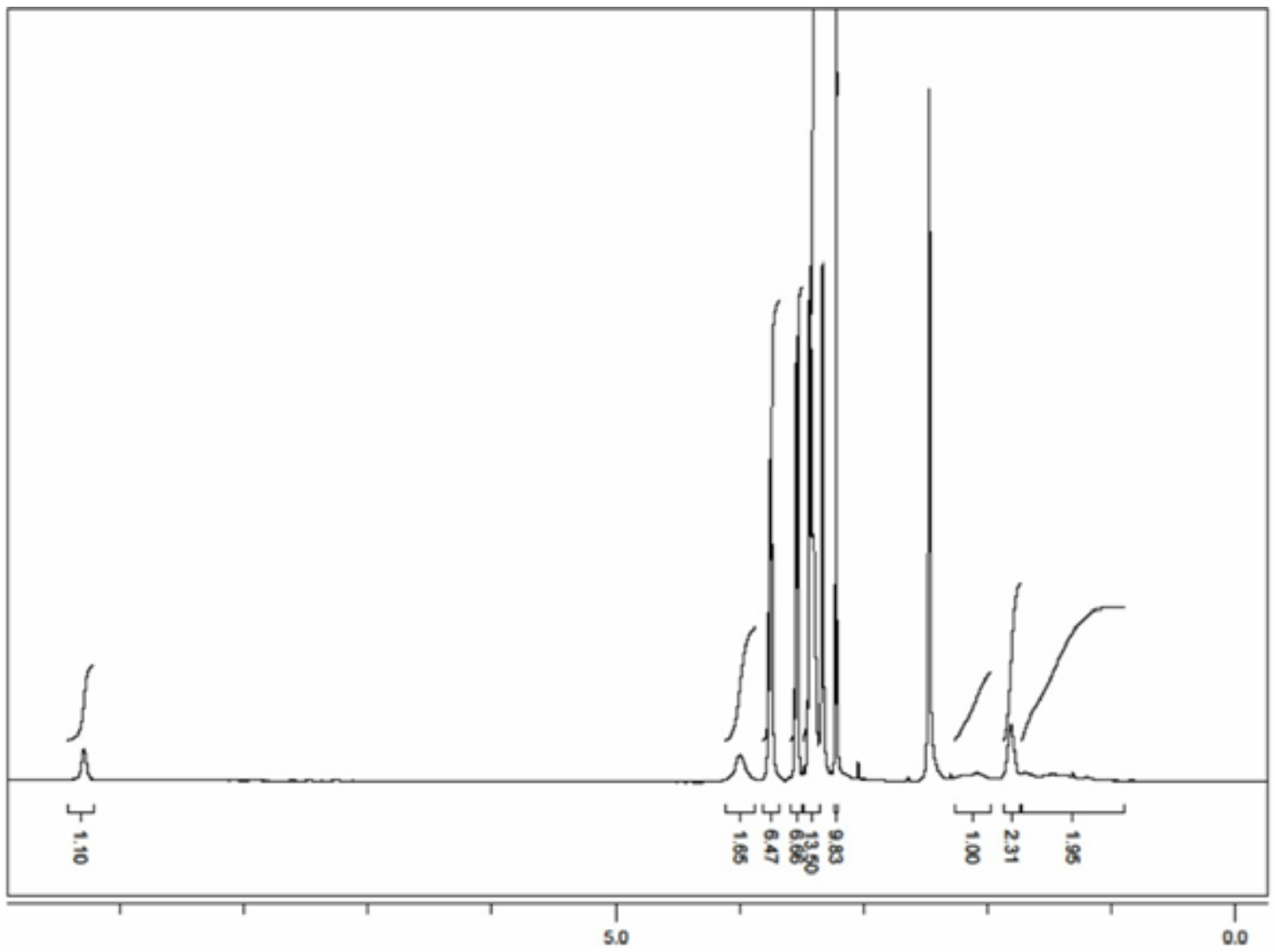

Figure 6: $^1$H NMR spectra of PAAPS with a molecular weight of 44200 g/mol.

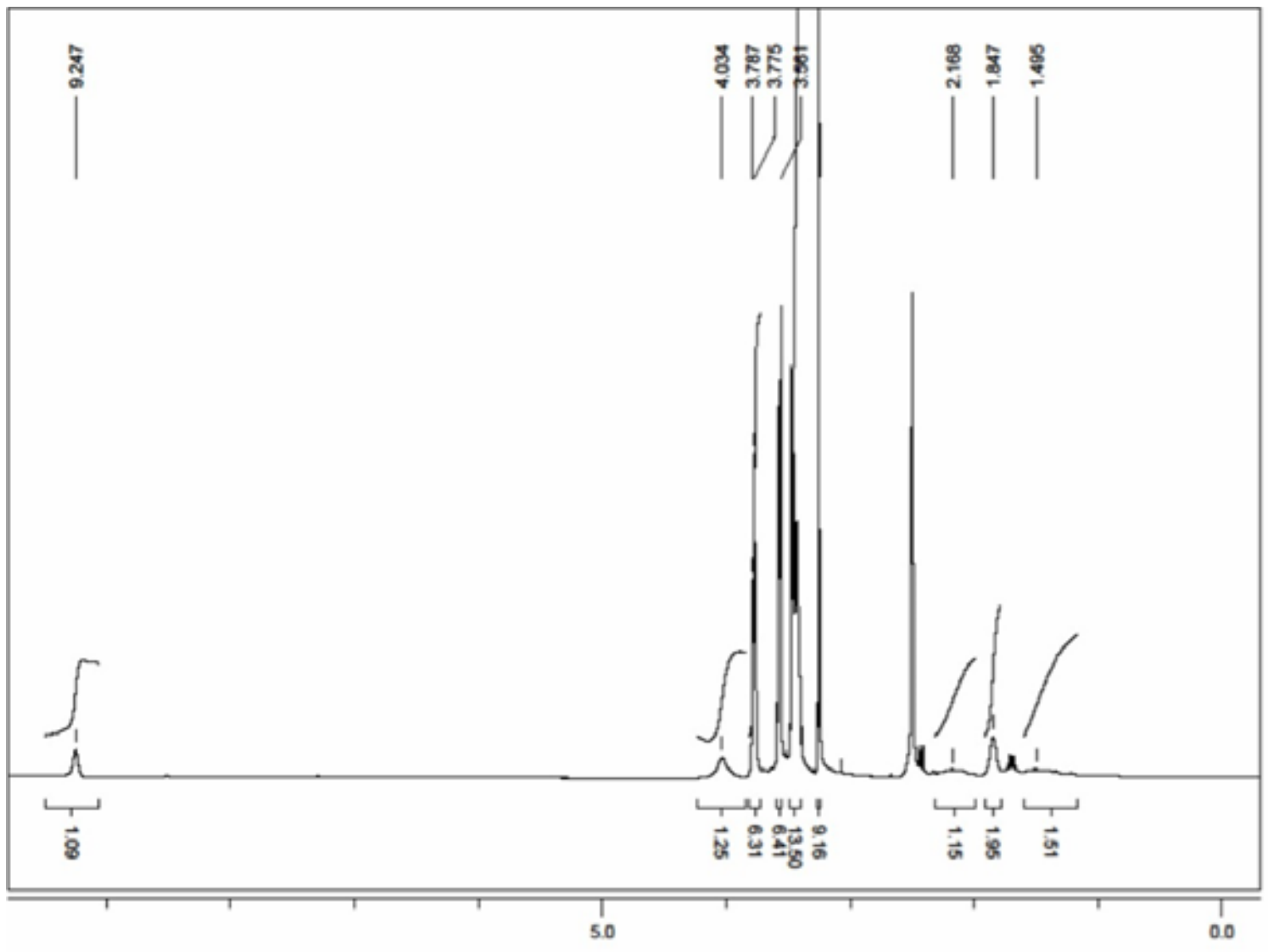

Figure 7: ¹H NMR spectra of PAAPS with a molecular weight of 51600 g/mol.

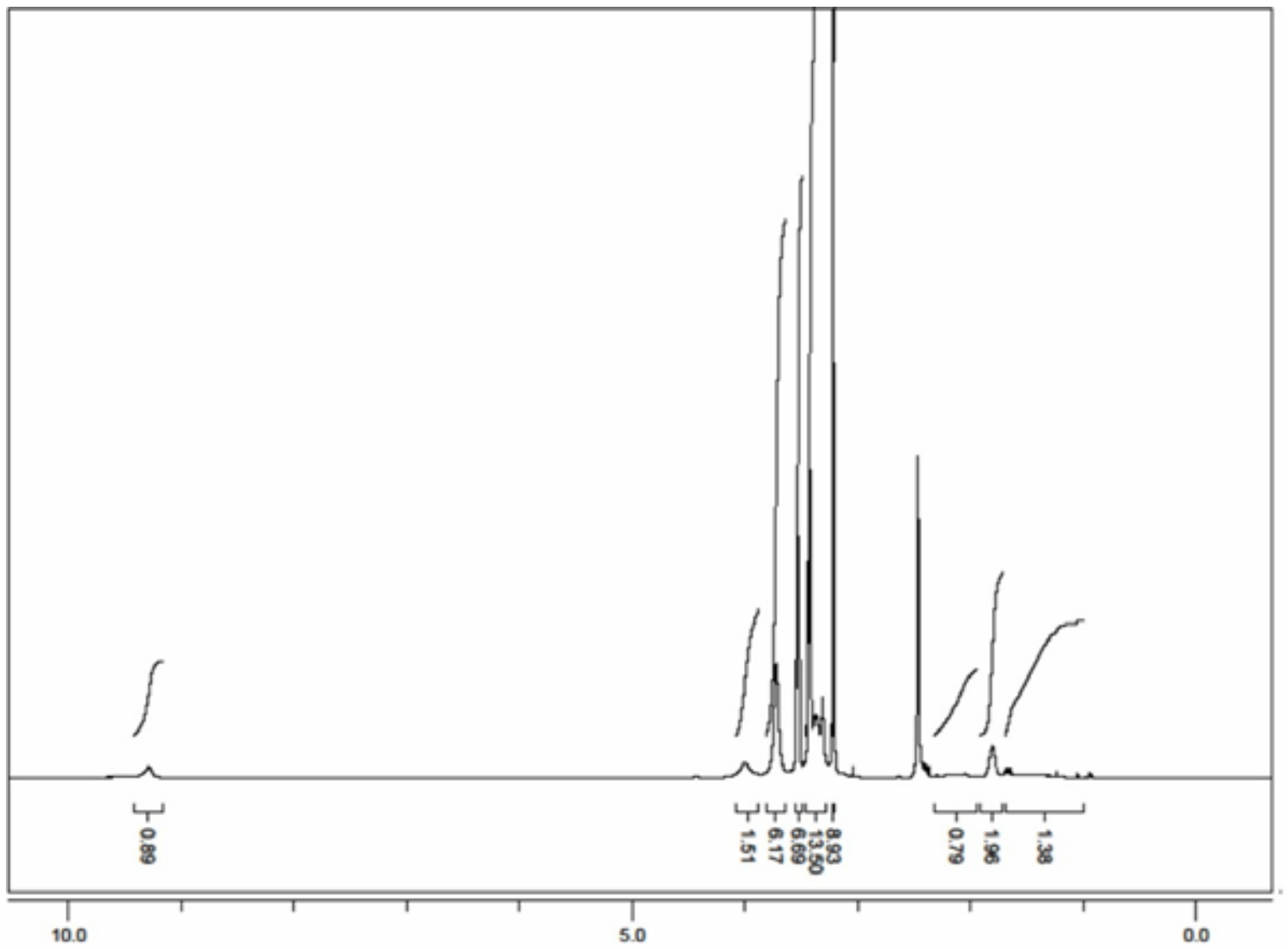

Figure 8: ¹H NMR spectra of PAAPS with a molecular weight of 99500 g/mol.

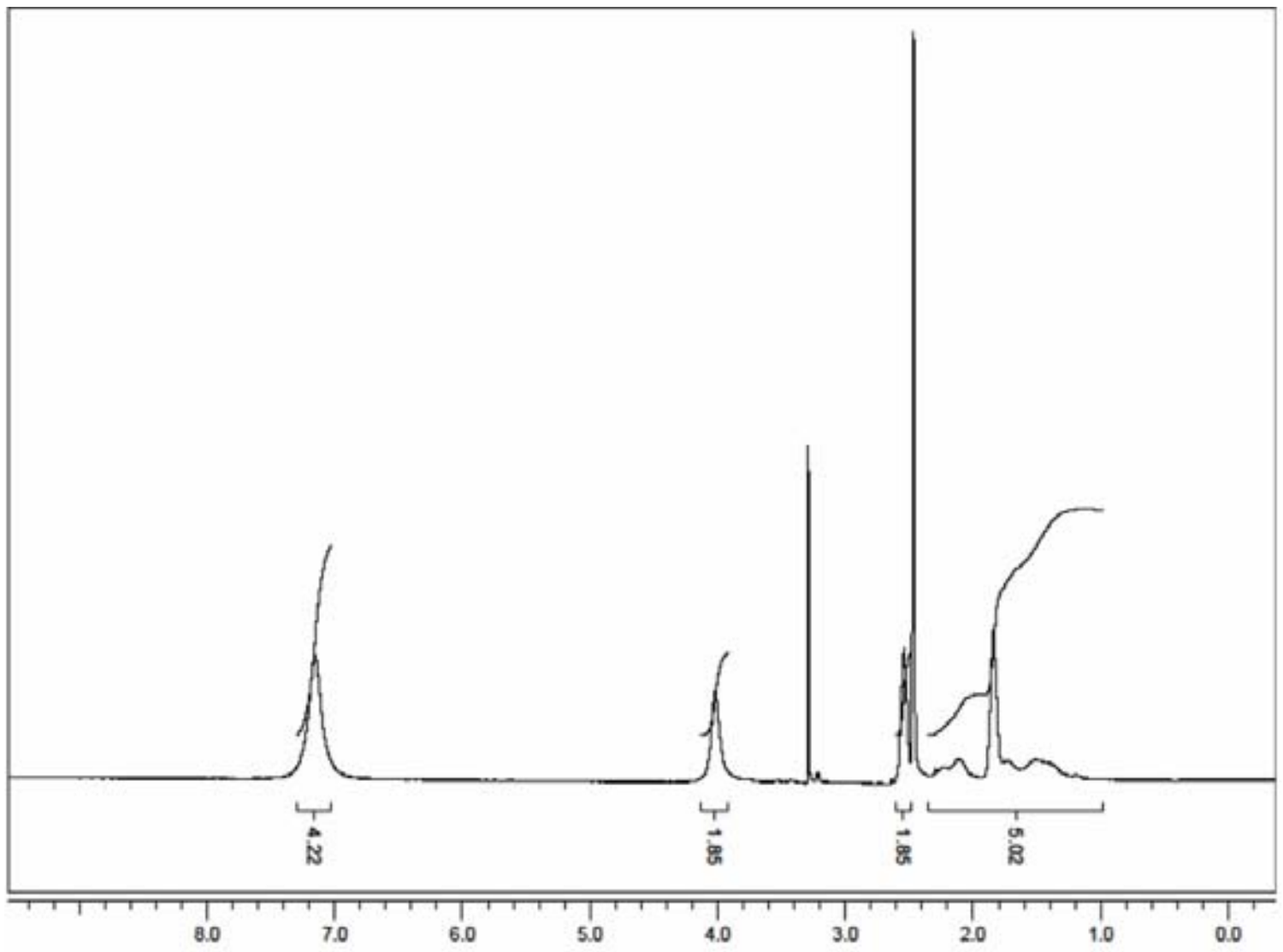

Figure 9: $^{1}$H NMR spectra of the modified molecule PAPS (PAAPS with $NH_4^+$ as cation).

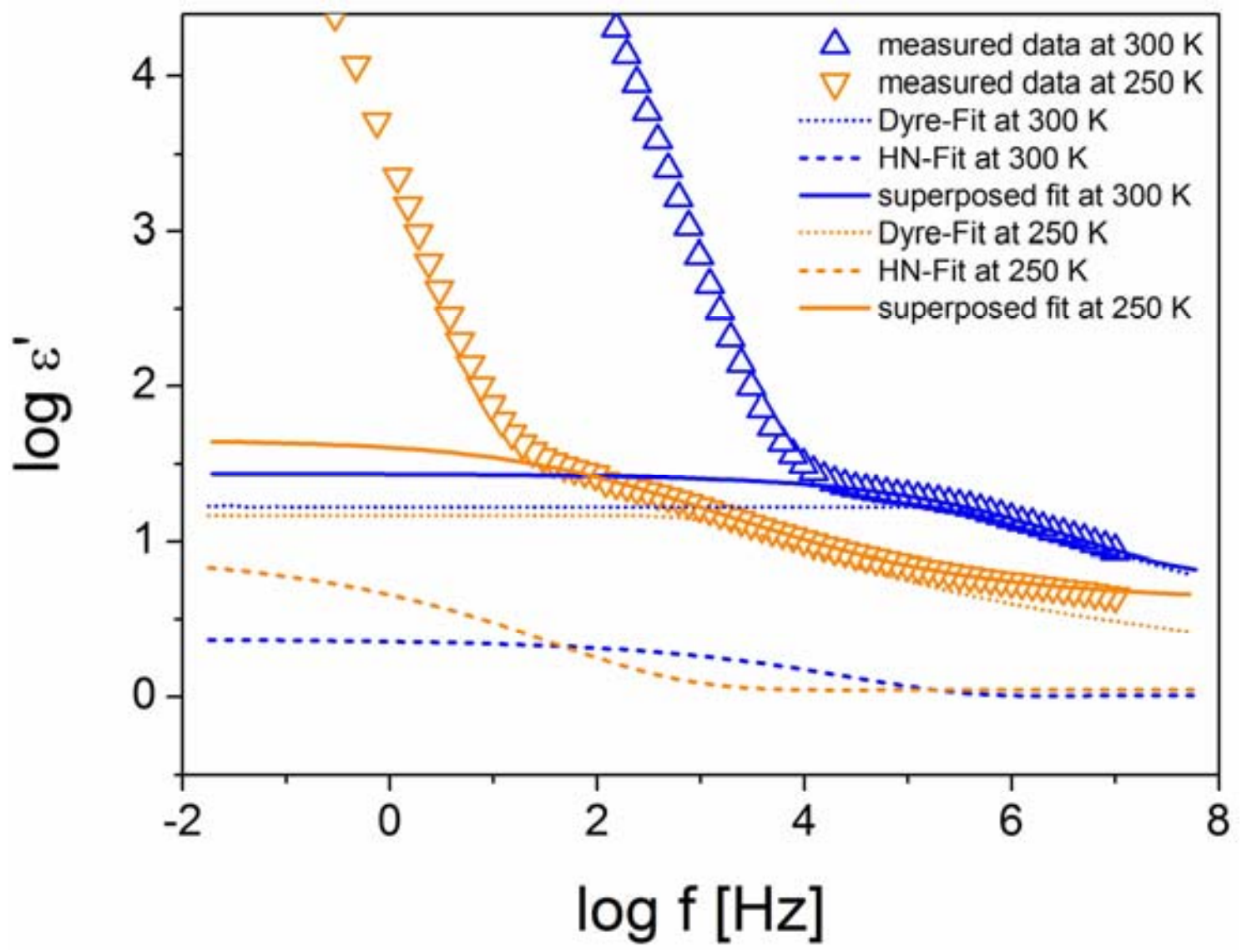

Figure 10: $\varepsilon^{I}$ *vs.* frequency of the PAAPS sample with a molecular weight Mn=99500 g/mol at two temperatures as indicated. The step represents the electrical energy storage of the bulk and is caused by a superposition of a molecular relaxation process fitted with a Havriliak-Negami (HN) function in $\varepsilon^{II}$ *vs.* frequency and a conductivity relaxation (that stores energy similar to a polaron) deduced by a Dyre-fit in $\sigma^{I}$ *vs.* frequency.

| notification | | Tg | log $\sigma_0$ at Tg + 50K | log $\sigma_0$ at Tg + 100K | chemical structure |
|---|---|---|---|---|---|
| IL | PIL | in [K] | in [S/cm] | in [S/cm] | |
| PAAPS, 9700 g/mol | --- | 225 | -6.65 ± 0.1 | -4.61 ± 0.1 | 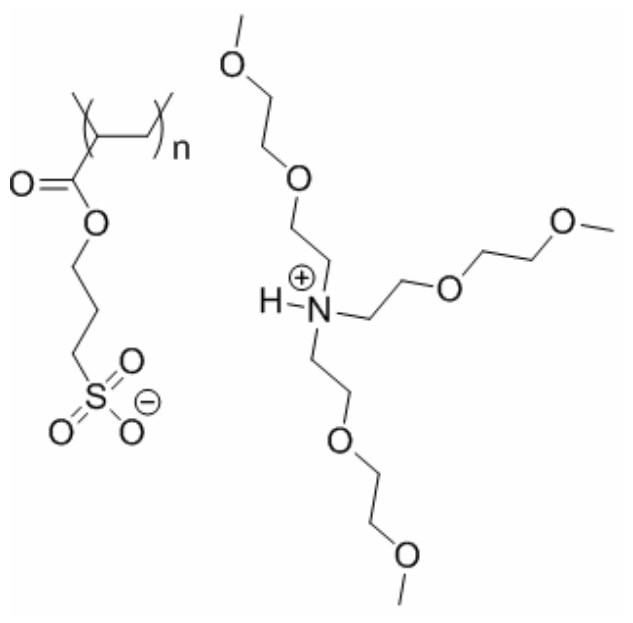 |
| --- | PAAPS, 99500 g/mol | 216 | -6.74 ± 0.1 | -4.69 ± 0.1 | |
| EVIM-NTf$_2$ | --- | 212 | -3.5 ± 0.1 | -2.3 ± 0.1 | 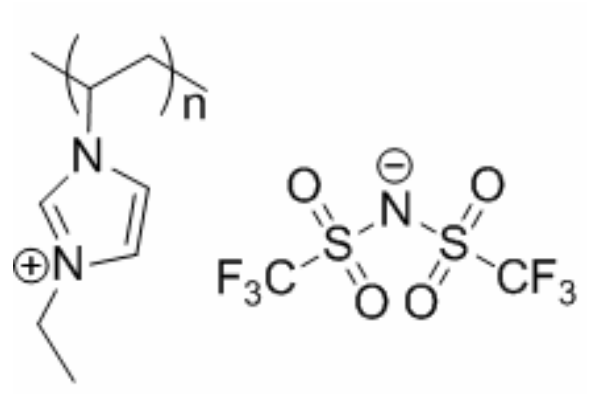 |
| --- | P EVIM-NTf$_2$ | 215 | -8.9 ± 0.1 | -5.5 ± 0.1 | |
| IL-1 | --- | 211 | -5.2 ± 0.1 | -2.9 ± 0.1 | 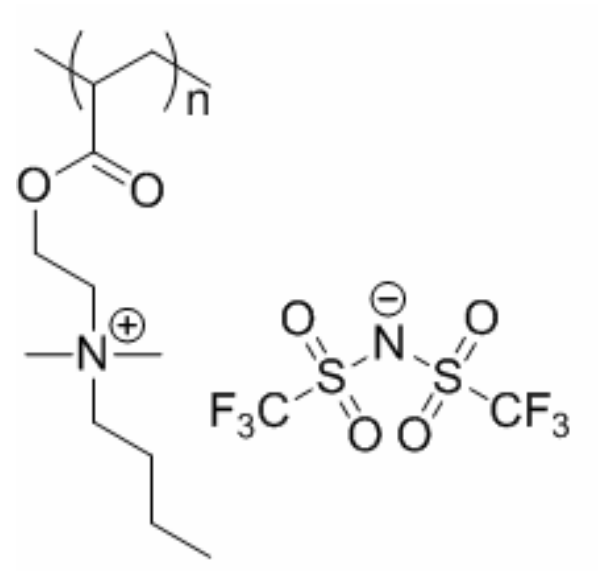 |
| --- | PolyIL-333 | 287 | -5.0 ± 0.1 | -3.5 ± 0.1 | |
| PVIM-NTf$_2$ | --- | 197 | -4.4 ± 0.1 | -2. 8 ± 0.1 | 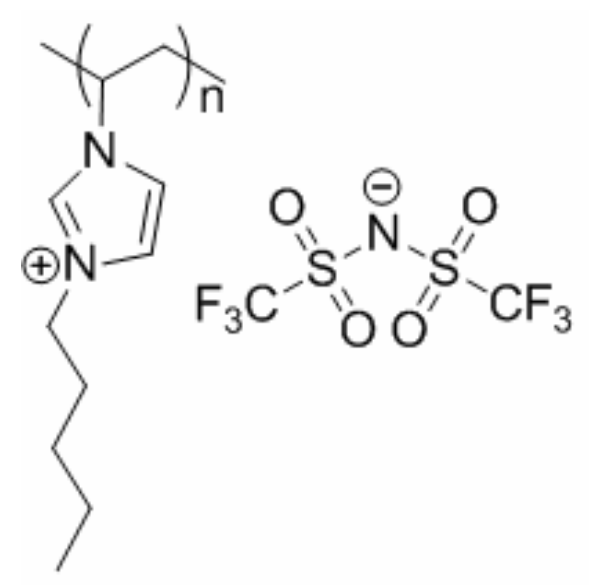 |
| | P PVIM-NTf$_2$ | 329 | -4.9 ± 0.1 | -3.4 ± 0.1 | |
| BVIM-NTf$_2$ | --- | 192 | -4.7 ± 0.1 | -2. 9 ± 0.1 | 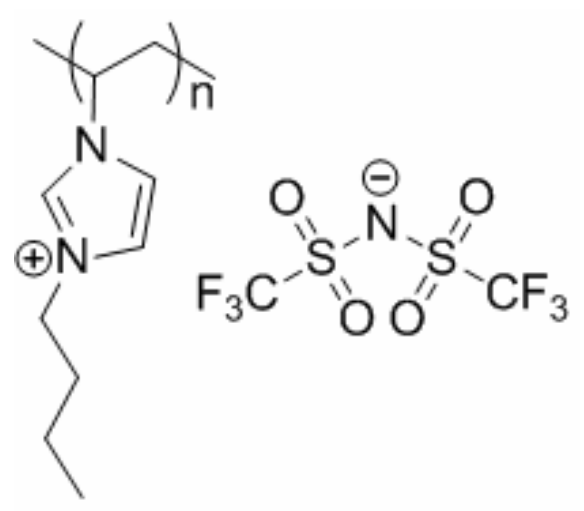 |
| --- | P BVIM-NTf$_2$ | 316 | -6.4 ± 0.1 | -4.9 ± 0.1 | |
| BPIMA-$PF_6$ | --- | 234 | -4.9 ± 0.1 | -3.7 ± 0.1 | 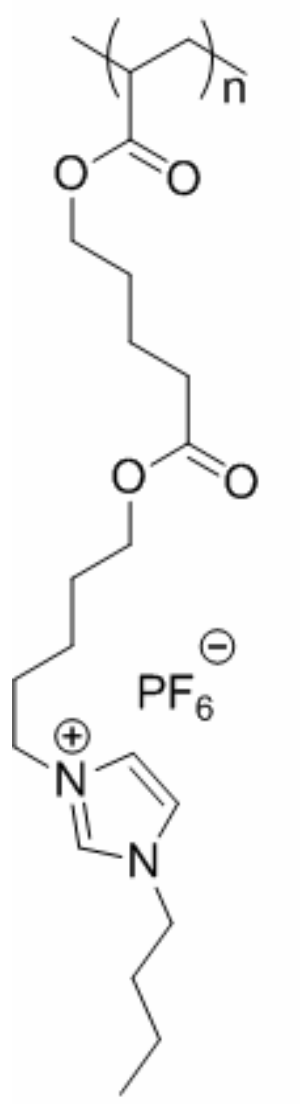 |
| --- | P BPIMA-$PF_6$ | 255 | -6.0 ± 0.1 | -4.0 ± 0.1 | |
| Py-IL | --- | 216 | -8.5 ± 0.1 | -4.8 ± 0.1 | 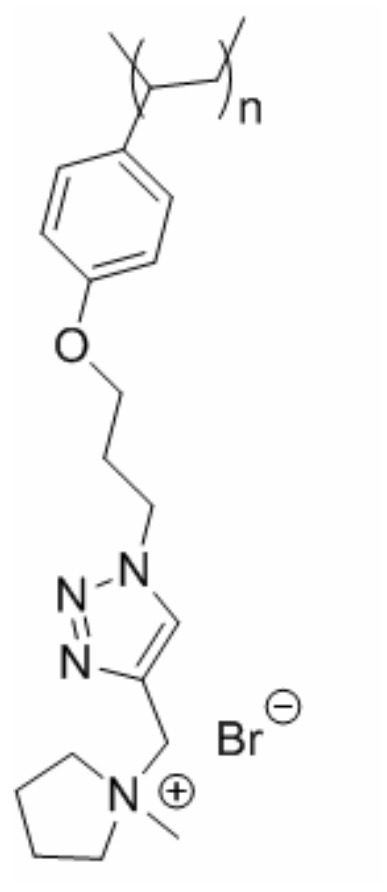 |
| --- | Py-Br | 207 | -17.0 | -14.2 | |

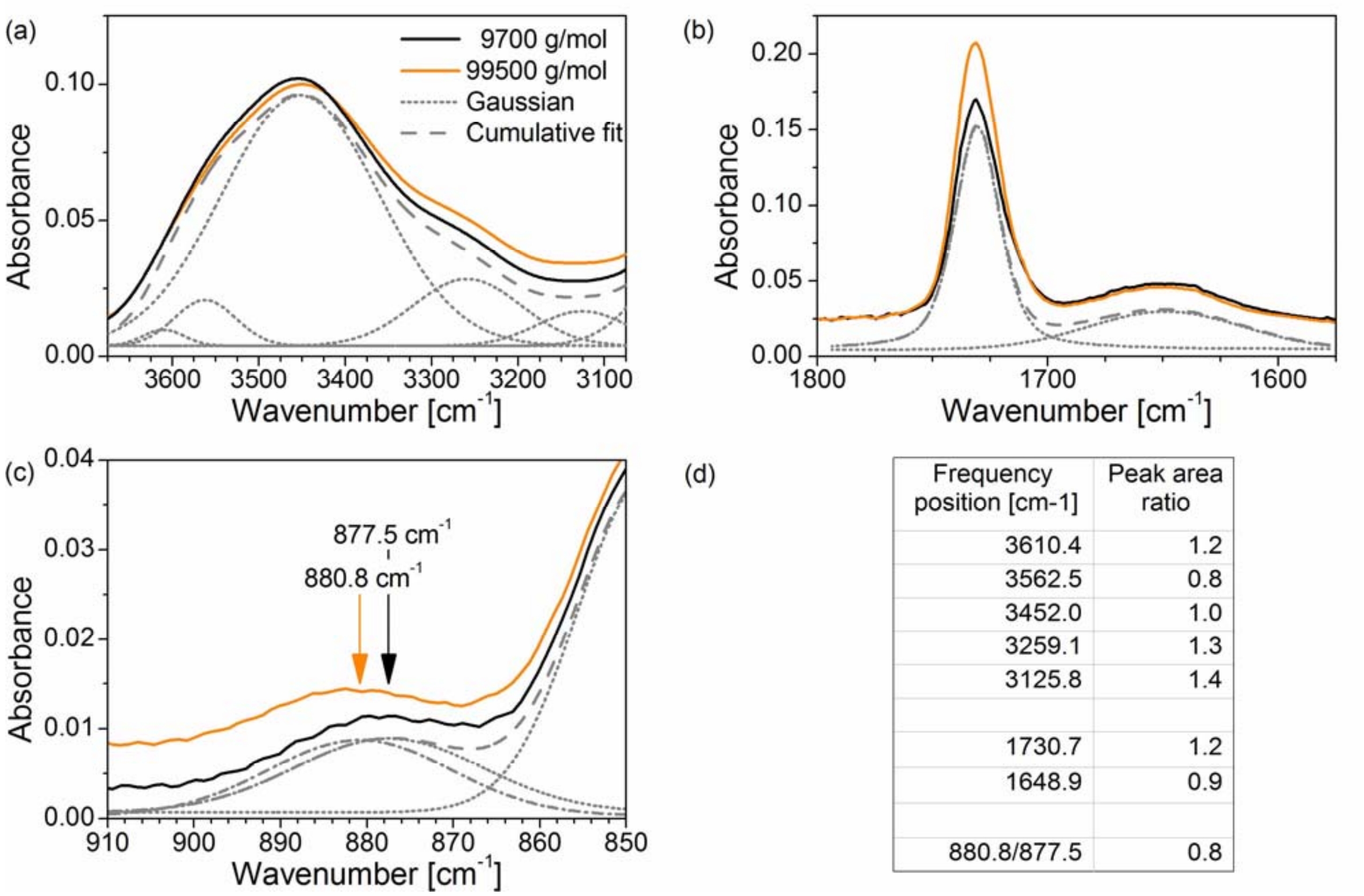

| Frequency position [cm-1] | Peak area ratio |
|---|---|
| 3610.4 | 1.2 |
| 3562.5 | 0.8 |
| 3452.0 | 1.0 |
| 3259.1 | 1.3 |
| 3125.8 | 1.4 |
| | |
| 1730.7 | 1.2 |
| 1648.9 | 0.9 |
| | |
| 880.8/877.5 | 0.8 |

Figure 11: FTIR spectra of the low molecular PAAPS (9700 g/mol) sample as well as the one with 99500 g/mol at selected frequency regions. The spectra are modeled by (a sum of) Gaussians (except for $\bar{\nu}$ = 1730.7 cm$^{-1}$, where a Voigt profile fits better), while the frequency positions and the peak widths are subsequently fixed.
(a) The N-H stretching vibration falls into the same region as the C-H stretching vibration and the C=O stretching overtone ([1]). In addition, this spectral range indicates hydrogen bonding. In general the frequency position of a stretching vibrations decreases when hydrogen bonds are formed, while their intensity rises and the peak becomes broader ([2]).Those criteria can be traced here. (b) The C=O stretching vibration of the ester group ($\bar{\nu}$ = 1730.7 cm$^{-1}$) and N-H bending vibration ($\bar{\nu}$ = 1648.9 cm$^{-1}$) appear at the same potions among the samples, but the intensity is pronounced for the highly polymeric system. (c) The N-H bending vibration for PAAPS with 995000 g/mol is shifted to higher wavenumbers compared to the one with 9700 g/mol ($\bar{\nu}$ = 880.8 cm$^{-1}$ *vs.* $\bar{\nu}$ = 877.5 cm$^{-1}$ ) and is less intense. In contrast to the stretching vibration (a), for bending vibrations the frequency rises with hydrogen bonding ([2]). Consequently, FTIR results indicate an enhanced hydrogen bonding in the highly polymeric system as compared to PAAPS with 9700 g/mol. Panel (d) gives the ratio of the area under the curves for the particular peaks, respectively.